%% file: articolo.tex
\documentclass[a4paper,11pt]{article}
\pdfoutput=1
\usepackage{jinstpub}
\usepackage[section]{placeins}
\usepackage{subfig}
\usepackage{multirow}
\usepackage{setspace}
\author{\hspace{-3pt}{\note[*]{Corresponding author.}}}
\author[a]{A.A.~Alves Jr,}
\author[b]{L.~Anderlini,}
\author[c]{M.~Anelli,}
\author[a,1]{R.~Antunes Nobrega,\note{Now at Universidade de
Juiz de Fora, Juiz de Fora, Brazil.}}
\author[{a,d}]{G.~Auriemma,}
\author[e]{W.~Baldini,}
\author[c]{G.~Bencivenni,}
\author[f,2]{R.~Berutti,\note{Now at CRS4 , Parco Scientifico e 
Tecnologico della Sardegna, Pula (Cagliari), Italy.}} 
\author[b]{A.~Bizzeti,}
\author[a]{V.~Bocci,}
\author[g]{N.~Bondar,}
\author[f]{W.~Bonivento,}
\author[g]{B.~Botchin,}
\author[f]{S.~Cadeddu,}
\author[c]{P.~Campana,}
\author[{h,i}]{G.~Carboni,}
\author[f]{A.~Cardini,}
\author[c]{M.~Carletti,}
\author[c]{P.~Ciambrone,}
\author[c]{E.~Dan\'e,}
\author[h]{S.~De~Capua,}
\author[f,3]{V.~De Leo,\note{Now at Linkalab, Complex Systems 
Computational Laboratory, Cagliari, Italy.}}
\author[f]{C.~Deplano,}
\author[c]{P.~De Simone,}
\author[{f,j}]{F.~Dettori,}
\author[k,l]{A.~Falabella,}
\author[m]{F.~Ferreira Rodriguez,}
\author[{b,n}]{M.~Frosini,}
\author[c,4]{S.~Furcas,\note{Now at Sezione INFN di Milano, Milano, Italy.}}
\author[a]{E.~Furfaro,}
\author[b]{G.~Graziani,}
\author[{o,p}]{L.~Gruber,}
\author[q]{G.~Haefeli,}
\author[g]{A.~Kashchuk,}
\author[a]{F.~Iacoangeli,}
\author[f]{A.~Lai,}
\author[c]{G.~Lanfranchi,}
\author[b]{M.~Lenzi,}
\author[g]{O.Levitskaya,}
\author[o]{K.~Mair,}
\author[g]{O.~Maev,}
\author[f]{G.~Manca,}
\author[f]{M.~Mara,}
\author[{a,*}]{G.~Martellotti,}
\author[r]{A.~Massafferri Rodrigues,}
\author[{h,i}]{R.~Messi,}
\author[c]{F.~Murtas,}
\author[g]{P.~Neustroev,}
\author[{f,j}]{R.G.C.~Oldeman,}
\author[c]{M.~Palutan,}
\author[b]{G.~Passaleva,}
\author[{a,s}]{G.~Penso,}
\author[a]{D.~Pinci,}
\author[m]{E.~Polycarpo,}
\author[f,5]{D.~Raspino,\note{Now at ISIS-STFC, Rutherford Appleton 
Laboratory, UK.}}
\author[{h,i}]{G.~Sabatino,}
\author[{f,j}]{B.~Saitta,}
\author[h]{A.~Salamon,}
\author[a]{R.~Santacesaria,}
\author[{h,i}]{E.~Santovetti,}
\author[c]{A.~Saputi,}
\author[{c,s}]{A.~Sarti,}
\author[{a,d}]{C.~Satriano,}
\author[h]{A.~Satta,}
\author[{e,l}]{M.~Savri\'e,}
\author[o]{B.~Schmidt,}
\author[o]{T.~Schneider,}
\author[c]{B.~Sciascia,}
\author[{a,s}]{A.~Sciubba,}
\author[f,6]{N.~Serra,\note{Now at Physik-Institut, 
Universit\"at Z\"urich, Z\"urich, Switzerland.}}
\author[t]{P.~Shatalov,}
\author[e]{S.~Vecchi,}
\author[{b,u}]{M.~Veltri,}
\author[g]{S.~Volkov,}
\author[g]{A.~Vorobyev}

\affiliation[a]{Sezione INFN di Roma, Roma, Italy}
\affiliation[b]{Sezione INFN di Firenze, Firenze, Italy}
\affiliation[c]{Laboratori Nazionali di Frascati dell'INFN, Frascati, Italy}
\affiliation[d]{Universit\`a della Basilicata, Potenza, Italy}
\affiliation[e]{Sezione INFN di Ferrara, Ferrara, Italy}
\affiliation[f]{Sezione INFN di Cagliari, Cagliari, Italy}
\affiliation[g]{Petersburg Nuclear Physics Institute, Gatchina,
St-Petersburg, Russia}
\affiliation[h]{Sezione INFN di Roma Tor Vergata, Roma, Italy}
\affiliation[i]{Universit\`a di Roma Tor Vergata, Roma, Italy}
\affiliation[j]{Universit\`a di Cagliari, Cagliari, Italy}
\affiliation[k]{Sezione INFN di Bologna, Bologna, Italy}
\affiliation[l]{Universit\`a di Ferrara, Ferrara, Italy}
\affiliation[m]{Instituto de F\'\i sica - Universidade Federal do
Rio de Janeiro (IF-UFRJ), Rio de Janeiro, Brazil}
\affiliation[n]{Universit\`a di Firenze, Firenze, Italy}
\affiliation[o]{European Organisation for Nuclear Research (CERN), 
Geneva, Switzerland}
\affiliation[p]{Technische Universit\"at Wien, Austria}
\affiliation[q]{Ecole Polytechnique F\'ed\'erale de Lausanne (EPFL), Lausanne, Switzerland}
\affiliation[r]{Centro Brasileiro de Pesquisas F\'isicas (CBPF),
Rio de Janeiro, Brazil}
\affiliation[s]{Sapienza, Universit\`a di Roma, Roma, Italy}
\affiliation[t]{ITEP, Moscow, Russia}
\affiliation[u]{Universit\`a di Urbino, Urbino, Italy}
\title{Performance of the LHCb muon system}
\emailAdd{giuseppe.martellotti@roma1.infn.it}

\abstract{
The performance of the LHCb Muon system and its stability  
across the full 2010 data taking 
with LHC running at $\sqrt{s}$ = 7~TeV energy is studied.
The optimization of the detector setting and the time calibration 
performed with
the first collisions delivered by LHC is described.    Particle rates, 
measured  for the wide range of luminosities and beam operation 
conditions experienced during the run, are 
compared with the values expected from simulation. 
The space and time alignment of the detectors, 
chamber efficiency, time resolution and cluster 
size are evaluated. The detector performance is found to be as expected from specifications or better. 
Notably the overall
efficiency is well above the design requirements. 
}


\renewcommand\afterKeywordsSpace{\thispagestyle{empty}\newpage}
\begin{document}
\maketitle\flushbottom
\singlespacing
\section{Introduction}
\label{sec:intro}

  LHCb is an experiment dedicated to heavy flavour physics at the LHC $pp$ collider. Its primary goal is  to look for  indirect evidence of new physics in CP violation and rare decays of beauty and charm hadrons. 

The LHCb apparatus~\cite{bib:lhcbpaper} is a single-arm 
forward spectrometer, consisting of  a series of sub-detectors aligned along the beam axis. 
A silicon-strip Vertex Locator (VELO) centered on the interaction point allows for precise vertex reconstruction. A dipole magnet with an integrated field of 4~Tm provides the bending for momentum measurement. Four multi-layer stations, one placed upstream and three downstream 
the magnet, ensure the tracking.  Silicon strips are used in the upstream station (TT) and in the downstream inner tracker (IT) while straw tubes are used in the downstream outer tracker (OT). 
Particle identification is provided by two ring imaging Cherenkov (RICH) detectors, by an electromagnetic and hadron calorimeter system (ECAL and HCAL) and by the Muon Detector. 

The calorimeters and the muon detector with their readout electronics are  designed to send information to the first level hardware trigger (L0) unambiguously identifying collision events of an LHC bunch crossing in a time window of 25~ns.
A second level software trigger (HLT) performs an almost complete event reconstruction using the information of the tracking detectors and selects specific channels of interest for the LHCb physics program~\cite{bib:trigger}.

The muon system consists of 5 detector stations with a total area of 435~$\mathrm{m}^{2}$, 1380 chambers of 20 different types for a total of 122k channels~\cite{bib:lhcbpaper}.
In 2009 a first setting of the detector was performed using cosmic rays~\cite{bib:cosmicpaper}. 
In 2010 the large data sample made available by the first LHC $pp$ collisions at $\sqrt{s}$~=~7~TeV, allowed for a rapid improvement of the detector working conditions and events were collected for a total of $\sim 37$~$\mathrm{pb}^{-1}$.

In this paper we describe the actions taken to optimize the detector 
performance, in particular the time calibration. We then report a detailed study
of the detector behaviour through the wide range of beam operation conditions spanned during the 2010 data taking.
Due to the continuous progress of the LHC, the luminosity spanned from $10^{27}$ to $1.5 \times 10^{32}$~s$^{-1}$ cm$^{-2}$ with an average number $\mu$ of interactions per bunch crossing increasing up to $\mu$$\sim 2$, while the nominal design values of LHCb were a luminosity of  $2 \times 10^{32}$~s$^{-1}$~cm$^{-2}$  and a $\mu$$\sim 0.4$.
Particle rates were measured across the five orders of magnitude of the experienced luminosities. 
The detector performance in terms of space and time alignment of the detectors, chamber efficiency, time resolution, noise level and cluster 
size was evaluated and compared with 
expectations.
The system behaviour resulted to be in agreement with the design specifications or better as in the case of the muon detection efficiency
in the required 25~ns time window.  

Since the end of the 2010 run period considered here, LHCb has taken almost 
two orders of magnitude more data, with even higher instantaneous luminosities.
The detector performance has remained very similar.

\input{LHCbmu}
\input{operability}

\input{datasample}

\input{basic}

\input{timing}
\input{align}

\input{efficiency}

\section{Conclusions}
\label{sec:conclusions}

The muon detector was successfully operated since the first year of LHC physics.
Its performance has been studied across the five orders of magnitude
of luminosity experienced during 2010 and compared with expectation. The whole system demonstrated an
excellent reliability and stability.  Detector requirements in terms of cluster size, time resolution and efficiency are fullfilled. 

During 2010 a small number of dead channels accounted for an overall inefficiency in the muon detection  $\lesssim$~1\%. Most of these channels were cured
 in the LHC shutdown after the 2010 data taking.
Thanks to a good monitoring system and maintenance work,
temporary failures occurring during the run, mainly HV trips of single chamber gaps, accounted for per mil effects in the efficiency. 
A careful setting of the chamber working point and a precise timing intercalibration allowed to
reach a muon detection efficiency, mainly determined by the chamber 
time resolution, well above the design requirement of 99\% in all the 5 muon stations.

\begin{acknowledgments}
We express our gratitude to our colleagues in the CERN accelerator
departments for the excellent performance of the LHC. We thank the technical
and administrative staff at the LHCb institutes.

We acknowledge support from CERN and from the National Agencies: CAPES,
CNPq, FAPERJ and FINEP (Brazil); NSFC (China); CNRS/IN2P3 and Region
Auvergne (France); BMBF, DFG, HGF and MPG (Germany); SFI (Ireland); INFN
(Italy); FOM and NWO (The Netherlands); SCSR (Poland); ANCS/IFA (Romania);
MinES, Rosatom, RFBR and NRC ``Kurchatov Institute" (Russia); MinECo,
XuntaGal and GENCAT (Spain); SNSF and SER (Switzerland); NAS (Ukraine); STFC
(United Kingdom); NSF (USA). We also acknowledge the support received from
the ERC under FP7.

The Tier1 computing centers are supported by IN2P3 (France), KIT and BMBF
(Germany), INFN (Italy), NWO and SURF (The Netherlands), CIEMAT, IFAE and
UAB (Spain), GridPP (United Kingdom). We are thankful for the computing
resources put at our disposal by Yandex LLC (Russia), as well as to the
communities behind the multiple open source software packages that we depend
on.
\end{acknowledgments}

\end{document}

%% file: LHCbmu.tex
\vspace*{5truemm}
\section{The LHCb muon system}
\label{sec:muondet}
  
The muon detector system is designed to send binary information to the data acquisition (DAQ) and to the hardware processors of the muon trigger (L0MU) which, together with the calorimeter trigger, constitutes the bulk of the first level trigger. 

\begin{figure}
  \centering
  \mbox{\includegraphics[width=1.0\textwidth]{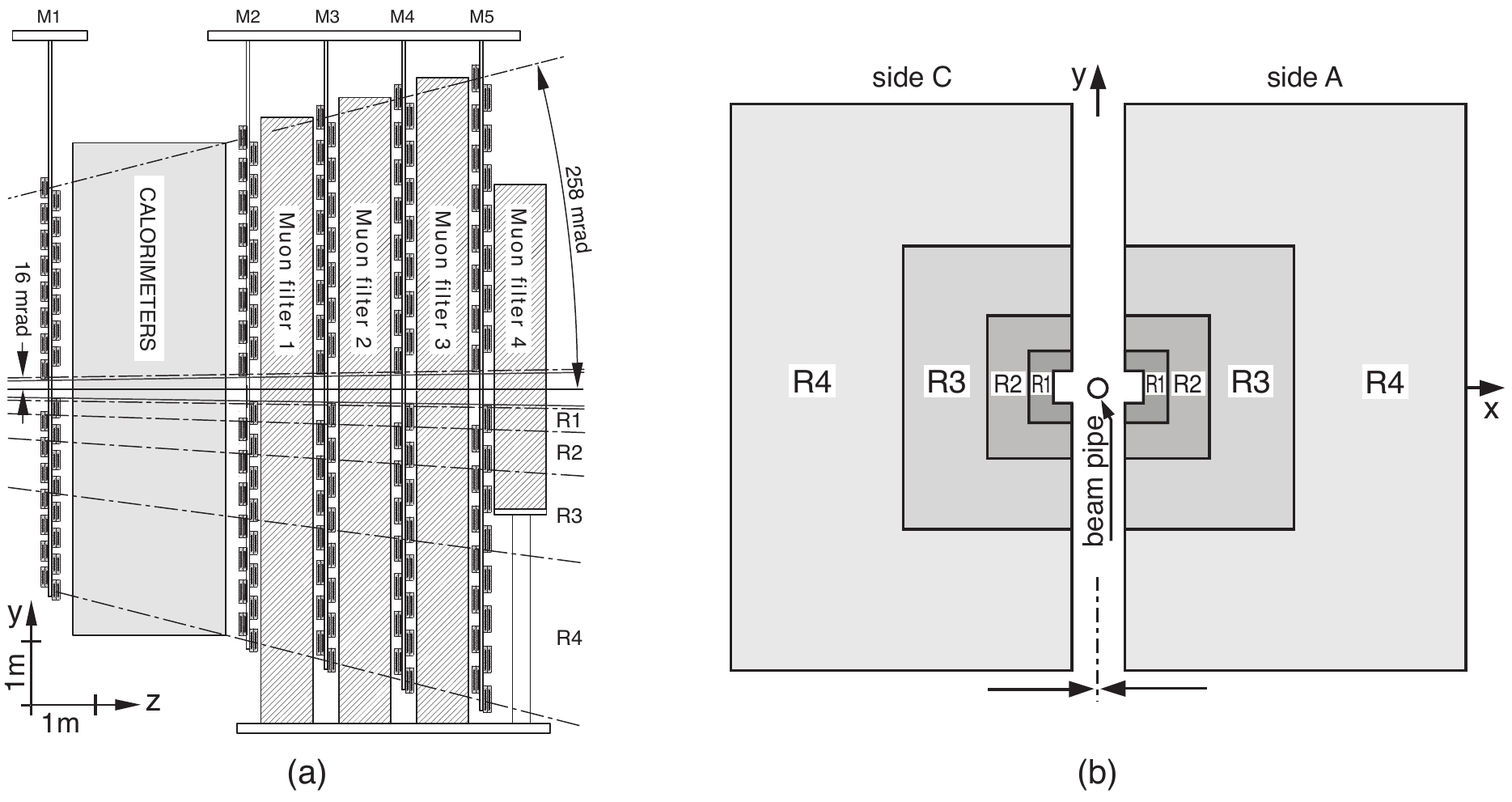}} 
  \caption{(a) Side view of the LHCb Muon Detector.
  (b) Station layout with the four regions R1--R4. }
\vspace*{6truemm}
  \label{fig:muondet}
\end{figure}

The detector  is composed of five stations (M1--M5) of rectangular shape, placed along the beam axis as shown in Fig.~\ref{fig:muondet}. 
Each station is equipped with 276 multi-wire proportional chambers (MWPCs) with the exception of the inner part of the first station, subject to the highest radiation, which is equipped with 12 GEM detectors~\cite{bib:GEM}. Each  station consists of two mechanically independent halves, called A and C sides, that can be horizontally moved to access the beam pipe and the detector chambers for installation and maintenance. Stations M2 to M5 are placed downstream the calorimeters and are interleaved with 80 cm thick iron absorbers.  Their information is used to identify and trace penetrating muons both in the online and offline analysis. Station M1 is instead located in front of the calorimeters and it is only used in the first level trigger.
 
The L0MU trigger processors~\cite{bib:L0MU} perform a stand-alone muon track reconstruction which requires to find hits in all the 5 stations and calculate the transverse momentum $p_{T}$ of the tracks. Muon candidates are accepted if their $p_{T}$ is above a given  threshold\footnote{The L0MU trigger is the logical ``or'' of a single-muon trigger with a $p_{T}$ threshold of 1.5 Gev/c and a di-muon trigger where a minimal value of 1.3 Gev/c is required for the geometrical mean of the first largest and the second largest muon $p_{T}$ found in the event.}. 
The hit alignment along a track is first verified in the four stations M2--M5 
searching for hits inside suitable fields of interest (FOI) projective to the interaction point. If this alignment is found, the hits of M2 and M3 stations are used to predict the track hit position in M1. If the M1 hit nearest to the prediction is found inside a suitable FOI, this hit and the one in M2 are used to define the track after the magnet deflection. The direction of such a track, its impact point at the magnet centre and the average $pp$ interaction point, provide a rough fast measurement of the magnet deflection and the $p_{T}$ used by L0MU.
The information of M1 station, placed in front of the calorimeter material, improves the $p_{T}$ resolution from $\sim$ 35~\% to $\sim$ 25~\%, with respect to what could be obtained using only the 4 downstream stations. The M1 information is however not helpful in the high level trigger or offline where a direct matching of the tracks reconstructed making use of the full spectrometer (T-tracks) with  the muon track segment detected in M2--M5 can be performed. The high resolution momentum of the matched T-track, typically ranging from 0.35 to 0.55~\%, is assigned to the muon. 

The geometry of the five stations is projective. The transverse dimensions of the stations scale with their distance from the interaction point.  The chambers are positioned to form, across the stations, adjacent projective towers pointing to the beam crossing position.

         The chambers are partitioned into \emph{physical channels} whose size is constrained by constructional reasons, or by requirements on their electrical capacitance and rate capability that influence the noise level and dead time of the front end (FE) electronics.
Appropriate combinations of physical channels are performed to build up  rectangular \emph{logical pads} having the $x$ and $y$ sizes required to obtain the desired performance of muon trigger and offline muon identification. 

\begin{figure}
  \centering
  \mbox{\includegraphics[width=1.0\textwidth]{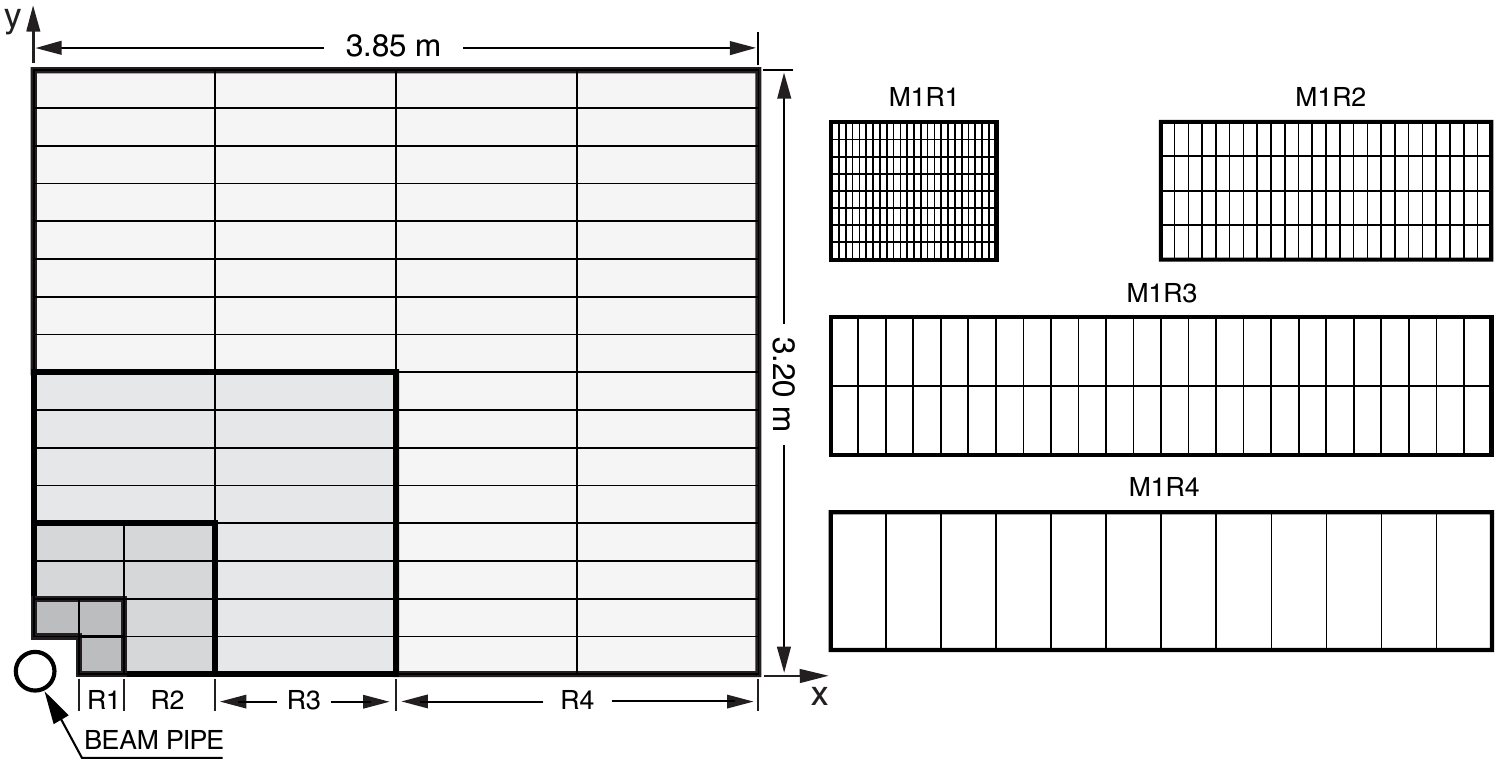}}    
  \caption{Left: a quadrant of M1 station. Each rectangle represents
  one chamber. Right: division into logical pads of four chambers belonging to the four regions of station M1. In stations M2, M3 (M4, M5)
the number of pad columns per chamber is double (half) the number in the corresponding region of station M1, while the number of pad rows is the same. }
\vspace*{6truemm}
  \label{fig:fig2}
\end{figure}

Each station is divided into
four regions with increasing distance from the beam axis as shown in  Fig.~\ref{fig:muondet}(b).
The linear dimensions of the
regions R1, R2, R3, R4, and the size of their logical pads, scale in 
the ratio 1:2:4:8 (see Fig.~\ref{fig:fig2}).        
Since the dipole magnet provides bending in the horizontal plane, the logical pad segmentation of muon chambers is finer in the horizontal direction $x$ than in the vertical direction $y$, to allow a good estimate of the momentum.  Stations M1, M2 and M3, used by the trigger to determine the track direction and the
$p_{T}$ of the candidate muon, have a higher $x$ granularity than stations M4 and M5, whose  main purpose is the identification of penetrating particles. 
In the inner region of the first station M1, the logical pad size is 1~cm in $x$ and 2.5~cm in $y$. In the other stations the vertical size $y$ just scales projectively with their distance from the interaction point; the $x$ granularity instead is two times finer in stations M2, M3 and two times larger in M4, M5.
The total number of logical pads is 55296.

Since the L0MU trigger requires 
a five-fold coincidence among all the stations, the
efficiency of each station must be $\ge$~99\%, within
a time window smaller than 25 ns, to obtain a trigger efficiency of at least 95\%. 
To comply with this stringent requirement, excellent time resolution and redundancy of the detector are needed.
The desired performance is obtained with an optimized charge-collection geometry and using a fast gas mixture\footnote{The gas mixture is \(\mathrm{Ar/CO_{2}/CF_{4}}\) $\simeq$ 40/55/5 for MWPCs and $\simeq$ 45/15/40 for GEM chambers.
}. Moreover the chambers are multi-gap detectors. In stations M2 to M5 the MWPCs consist of two coupled bi-gap detectors with two independent readouts.  
 In station M1, R2 to R4  the MWPC's have only two gas gaps with independent readout to minimize the material in front of the electromagnetic calorimeter. In
region M1R1 two superimposed triple-GEM chambers are used. In all cases, in standard running conditions, the two independent readout 
layers are connected to a logical ``or'' in the FE.

Since constructional constraints, as well as requirements on spatial resolution and rate capability,  strongly vary in different stations and regions of the detector, different readout techniques  were employed.
   In the high granularity regions M2/3 R1/2 
a double readout was adopted for the chambers: the physical channels are narrow vertical anode strips defining the $x$ resolution  and  larger cathode pads
 defining the $y$ resolution. Both signals are readout via FE channels and directly sent to the trigger and DAQ through optical links.
Logical pads are then obtained as a logical ``and'' between anode and cathode pads. 
In all the other stations and regions a single readout was adopted: the chambers are segmented into anode or into cathode pads generally smaller than the required logical pad. These pads are subsequently combined to build larger \emph{logical channels} that are sent to the trigger and DAQ.
In the large low resolution external regions R4, anode pads  are formed by soldering an appropriate number of adjacent wires. In the other regions, both in MWPCs and in GEM chambers, cathode pads are obtained with a segmented printed circuit board. 
A total of 122112 physical channels enter the front end electronics.

The FE boards house two custom made 8-channels ASICs called CARIOCA~\cite{bib:carioca} that can process both the negative and positive polarity signals from wires and cathode pads. Each channel  consists of a fast low-impedance charge-sensitive preamplifier, a main amplifier-shaper with base-line restorer and single-threshold discriminator.  The response  has a peaking time of 10~ns and a width not exceeding 60~ns to minimize the dead time.		
The CARIOCA output channels are routed to  
flexible logical units performing the logical ``or'' of a variable number of channels.  
Up to four adjacent physical pads are connected to build a logical pad.  In the M1 station, where the channel occupancy is high, the signals from the logical
pads are directly sent to the trigger and DAQ. In most of the other low occupancy regions, M2/3 R3/4 and M4/5 R2/3/4, several contiguous logical pads are further connected to build larger logical channels in the form of vertical and horizontal strips, with the aim of reducing the number of optical links to the trigger and DAQ. 
   The logical ``or'' is fully performed by logic units (DIALOG~\cite{bib:dialog1}) sitting on the FE boards only in part of the detector, while is completed on special intermediate boards in regions where the logical channel spans more than one FE board.
    
 A total of 25920 logical channels are built and finally routed to the Off Detector Electronics (ODE)~\cite{bib:ode}, where signals are tagged with the identification number of the bunch crossing (BXID) and sent to the
 trigger processors via optical links without zero suppression. 
  In the ODE boards the fine time information inside the 25~ns gate, 
measured by a 4-bit TDC ASIC~\cite{bib:dialog}, is added to the
data to be sent to the DAQ. The trasmission from the ODE to the 
acquisition system goes through the TELL1 boards~\cite{bib:TELL1} 
where the data are zero suppressed and suitably packed.

A distributed system, based on a CAN-bus network and 600 microcontrollers accessing single physical channels, performs the setting of the operating conditions and the monitoring of the detector~\cite{bib:bocci}

%% file: operability.tex
\section{Detector setting}
\label{sec:setting}

The operating conditions of the chambers were optimized through a procedure 
described
elsewhere~\cite{bib:optanatoli,bib:thranatoli} which
aims at reaching a
low and stable noise rate 
while  minimizing
ageing effects and cluster size. These conditions were satisfied by setting the thresholds to 6 noise r.m.s. and by choosing the minimal high voltage 
value allowing to obtain the required 99\% efficiency.

Although all the MWPCs have the same gas gap of 5~mm, different HV (from 2.53 to 2.65~kV) were required in different regions, depending on the pad size 
and the readout technique.
The chamber setting was kept constant during the whole 2010 run, except
for region M5R2 were voltage was reduced by 90~V when a large increase 
in luminosity produced a jump of the HV trip rate.
Thresholds were
lowered correspondingly to maintain high efficiency at the price of a slight increase of the noise
level.

The triple GEM detectors of M1R1 were operated with voltages of
435/425/415 V for the first part of the run. 
Voltages were later reduced twice by 5 V per gap for safer operations following
the increasing luminosity. 
\vspace*{-2truemm}

\section{Detector Operation performance in 2010}
\label{sec:functioning}

During 2010 the detector was operated to acquire some special calibration runs and many physics runs with $pp$ collisions events at $\sqrt{s}$ = 7~TeV. 
Due to the continuous progress of the LHC, the 2010
data span a wide range of luminosity values, from $10^{27}$ to 
$1.5 \times 10^{32}$ s$^{-1}$ cm$^{-2}$. The machine operations were accordingly evolving with time, from
the single colliding bunch per orbit of the first runs, to the 150 ns
spaced bunch trains of the highest intensity runs with 344 colliding bunches per orbit, equivalent to a 3.9 MHz collision rate, and the
first tests with 50~ns spaced bunches. Overall an integrated luminosity of $\sim 37$~$\mathrm{pb}^{-1}$ was collected. 

\begin{figure}
  \centering
  \mbox{\includegraphics[width=1\textwidth]{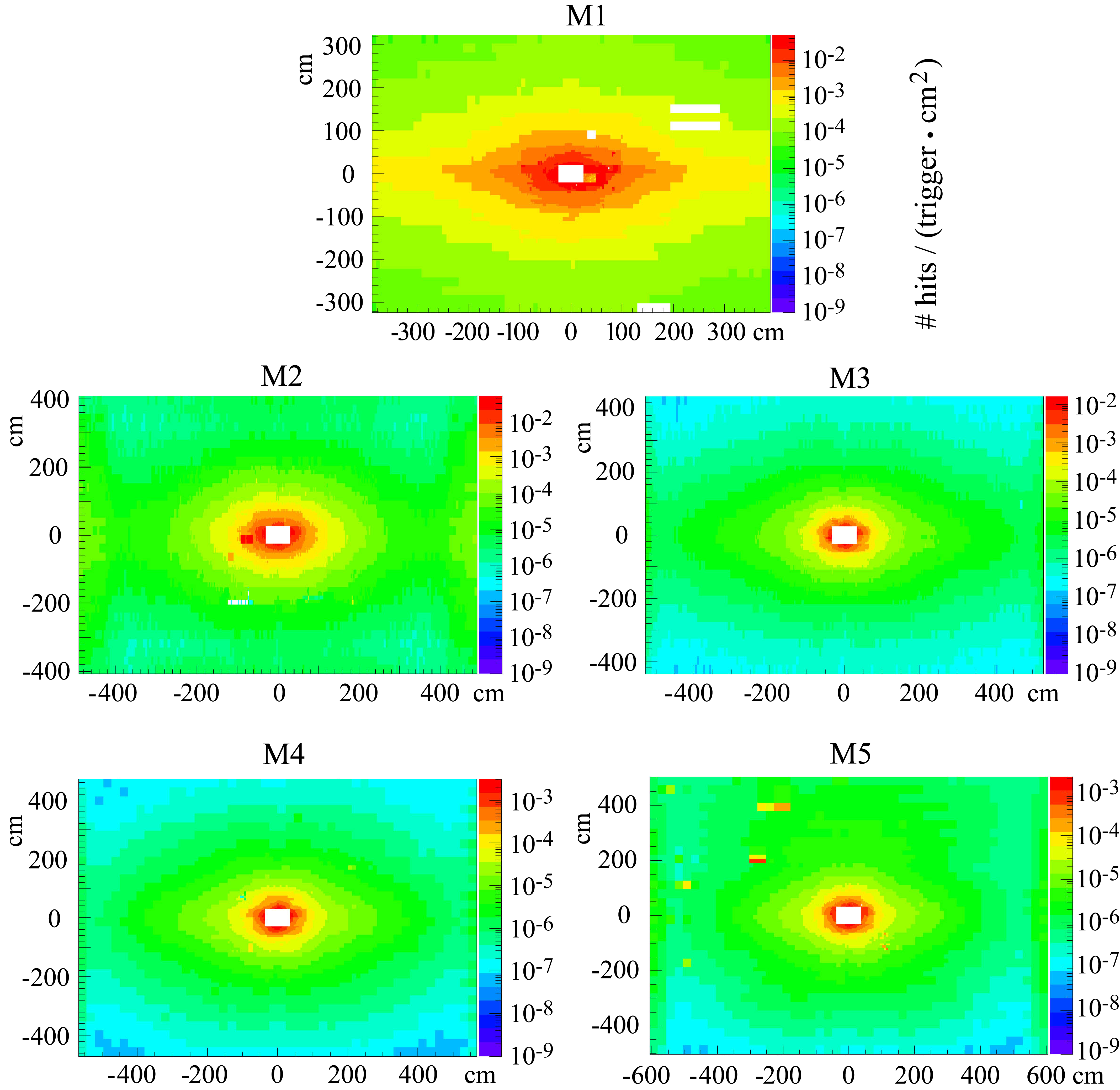}}
  \caption{Illumination map of the five detector stations in a
    typical 2010 physics run. The log color scales give the average number of hits per
    cm$^2$ per trigger for all the 55296 logical pads. Faulty channels
    giving no hits and a few noisy spots can be noticed.}
\vspace*{5truemm}  
\label{fig:padmap}
\end{figure}

As a first illustration of the muon detector operation performance, we show in Fig.~\ref{fig:padmap} a typical hit map of the five
stations during a physics run. The few holes present on the maps
correspond to dead channels. Most of them were due to hardware problems
in the readout chain known since the beginning of the data
taking\footnote{Almost all of them have been cured in the LHC shutdown after the 2010 data taking.}.
They affected only 129 of the 55296 logical pads (0.2\%). 
The detector failures occurred during the run, were mostly  
due to HV trips of single detector gaps and were recovered after some 
conditioning procedure.
 
The overall effect of dead channels on the muon tracking efficiency
over the full 2010 run
has been estimated~\cite{bib:performance-note} by counting the 
fraction of muon tracks 
with momentum larger than 6 GeV/c crossing one of the dead 
channels in Monte Carlo minimum bias events. It resulted to be 
below 1\%. 
For the stations M2--M5, due to the large redundancy given by the four 
independent gaps
per chamber, the effect of gap failures on the overall  
system efficiency was negligible ($<$ 0.1\%).
In the  case of M1, gap failures are more dangerous since there 
are only 2 gaps per chamber.
However, as it has been explained in section~\ref{sec:muondet}, this station is only used by the L0MU trigger and not
by the high level trigger and  offline reconstruction.   
If a detector of M1 is faulty, the L0MU trigger can be set to ignore the information of the trigger sector\footnote{A Trigger Sector is the 
small detector zone which is the basic unit where L0MU operates. 
In the case of M1 it directly contains the logical pads. In most of the detector it contains horizontal and vertical strips and the fired logical pads are identified by crossing the $x$ and $y$ strips.} 
containing the M1 dead channels and calculate $p_{T}$ with M2 and M3 hits. This results in a poorer resolution of the 
$p_{T}$ measurement and consequently a slight degradation of the trigger performance in the small affected region.
This procedure was used only in one case for a chamber in M1R1, 
when both its triple GEM detectors became very inefficient.
This resulted in  a negligible trigger efficiency loss at the Pt threshold, while this dead zone would have otherwise generated an overall 
inefficiency of the order of 3\% in the stand alone trigger tracking.

The noise in the detector should be below 1~kHz per physical 
channel in order to have a negligible rate  (below 1~Hz) 
of fake muon triggers~\cite{bib:l0mudesign}. 
Even lower rates are desirable to suppress the noise 
contribution to muon misidentification in the
offline reconstruction.
The level of detector noise was checked regularly with dedicated
runs of random triggers in absence of beam and 
the  noise rate per physical channel was computed for 
each detector region from the
multiplicity of firing logical channels or logical pads. 
The fraction of channels having a noise rate larger than 10 kHz 
is typically lower than 0.1\%. Such values are well below the
noise level causing a measurable rate of accidental triggers.

%% file: datasample.tex
\vspace*{6truemm}
\section{Data samples and track reconstruction}
\label{sec:datasample}

In order to test the muon detector response in the wide range of 
conditions experienced during 2010, an appropriate set of sample runs over the full period  
was chosen and analysed. 

For time calibration and time performance study, some special runs were taken:
\begin{itemize}
\item Events for time alignment (TAE) were recorded in a larger time gate of 125~ns, instead of the standard 25~ns, around the triggered bunch crossing.
\end{itemize} 

For the measurement of particle rates, cluster size and time resolution,
events triggered by some minimum
bias condition, independently of the muon detector response, were used:
\begin{itemize}
\item Minimum bias trigger (L0MB), requiring the total energy released in
  the HCAL to be more than 320 MeV;
\item ``Microbias'' single track trigger (microbias),
  requiring some hits compatible with a track in the VELO or first
  tracking stations; 
\end{itemize}

 For spatial alignment studies (section \ref{sec:align}) and detector efficiency measurements (section \ref{sec:effi}), when the statistics of muon tracks in the minimum bias samples was not adequate, events collected
with standard physics triggers were used. The samples chosen and the procedures adopted to avoid the bias introduced by the trigger will be described there.

 Monte Carlo simulation was used to compare the observed detector performance with the expectation, and to verify the analysis procedures.
Standard samples of LHCb Monte Carlo events
simulating minimum bias 
$pp$ interactions and production of prompt $J/\psi$ decaying to
$\mu^+ \mu^-$ were used. Other special samples used for particular needs will be
described on the next sections. The events were generated using PYTHIA 
6.4~\cite{pythia} 
to describe the $pp$ collisions and GEANT4~\cite{geant} for the 
LHCb detector simulation. 
\hfill \\

Regarding track reconstruction, in the present analysis, 
standalone muon tracks  (M-tracks) 
are reconstructed with an algorithm similar to the one used in the muon
high level trigger~\cite{hlt}.
Tracks are reconstructed starting from the firing logical
pads. If adjacent fired pads are found, they are clusterized, both in $x$ and $y$, to obtain track hits. 
Clusters are allowed to extend across adjacent chambers even belonging to different regions. The $x$ and $y$ of the cluster
barycenter and the $z$ of the station midplane are assumed as the track hit coordinates. 
Hits aligned with the average position of the $pp$
collision point are selected by a combinatorial algorithm.
Track hits are fitted to a straight line and 
quality cuts are applied depending on the measured quantities.
M-tracks are usually reconstructed using the information of all the stations unless required by the specific analysis.

M-tracks are required to match with one track reconstructed in the tracking detectors (T-tracks) only when this is needed to reduce  background or to have a good muon momentum measurement. The matching is performed on the basis of the comparison, in one or more muon stations, of the M-track hit coordinates with the T-track extrapolations at the corresponding $z$ positions. The compatibility of the M-track and 
T-track slopes is also required. This procedure is similar to the one used in HLT and offline. In all cases the muon momentum is measured from the matched T-track, forgetting the poor resolution momentum that could be extracted from the M-track as in the L0MU trigger.

%% file: basic.tex
\section{Rates}
\label{sec:rates}

The rate capability was one of the key request for the choice of the
technology and the design of the muon detector. Detailed simulations were
developed~\cite{bib:romarates} to evaluate the particle rates and
radiation doses expected for the nominal LHCb operations at an energy of $\sqrt{s}$ = 14 TeV and a
luminosity of $2 \times 10^{32}$~s$^{-1}$~cm$^{-2}$. With respect to the radiation hardness, the detector was designed
to stand, for the 10 years of
planned LHCb operation, rates larger than a factor 3 in stations M2--M5 and a factor 2
in station M1 as compared with these simulations.
The main reasons for the large safety factors were the uncertainies
on the track multiplicity in pp collisions predicted by PYTHIA~\cite{pythia}, and the uncertainties on the simulation of low energy particles from showers around the beam pipe and particle backscattering from the material surrounding the detector. 

From 2010 data we measured the actual rates at $\sqrt{s}$ = 7~TeV 
for a wide range of
luminosities and compared them with expectations~\cite{bib:performance-note}. 
From the number of hits seen by triggered events we can extract  the average number of hits per visible interaction and evaluate the contribution due to the current interactions and the one not due to the triggered collision. The first contribution is by definition independent of  luminosity. The second one is expected to be dependent on the beam operation conditions. It can
be due to beam background, residual detector noise or, more importantly, to late hits originating from collisions in previous bunches (\emph{spillover}). 
Particle back scattering from the cavern or from heavy objects surrounding the beam pipe is expected to give a significant contribution to spillover. 
Also detector effects like late cross-talk or afterpulses can contribute.

Let's consider the firing logical pads, also called hits in this study, no matter if they clusterize or not. 
Given an unbiased trigger $T$, the number of hits per unit surface and per triggered event
\begin{equation}
\label{eq:rpt}
r_T=dN_h/dSdN_T  
\end{equation}
 was computed for each chamber, and the average number of hits in
each detector region was calculated after removing the few chambers with channels affected by some pathology.

\begin{figure}[t]
  \centering
  \mbox{\includegraphics[width=.9\textwidth]{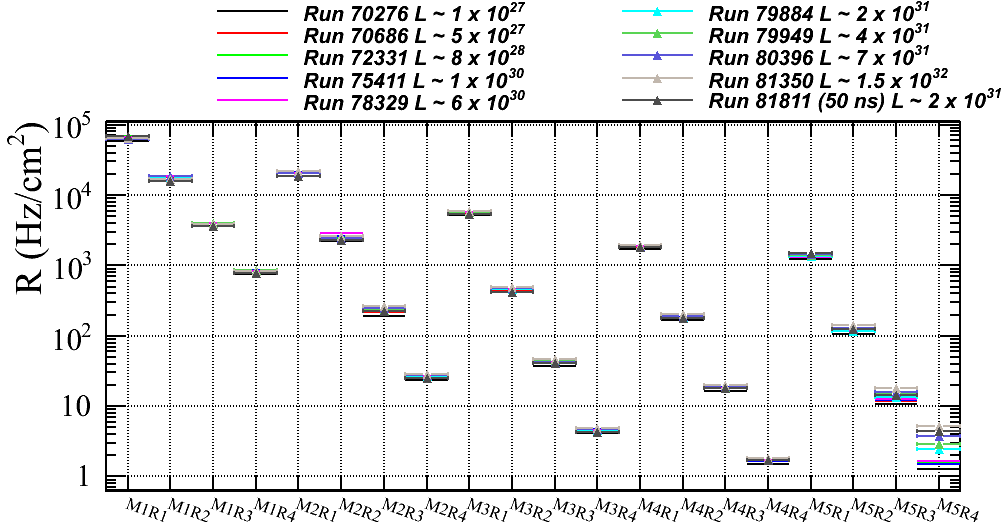}}
\caption{Normalized rates $R$, extrapolated to the nominal luminosity (equation \ref{eq:norm}), in
    the 20 muon detector regions, for the 2010 sample runs acquired in different beam operation conditions. The rates refer to the 
    in-time hits (within the trigger 25 ns gate).}
\vspace*{0truemm}
\label{fig:rate}
\end{figure}

The microbias trigger provided 
the least biased trigger suitable for the measurement. 
For each sample, the average number $\mu$ of interactions visible in the LHCb detector 
per beam crossing was evaluated from the fraction $f_0$ of beam crossing events not producing a trigger 
\begin{equation}
\label{eq:mu}
  f_0 = P(0;\mu) = e^{-\mu}
\end{equation}
where $P$ is the Poisson distribution. The pile-up factor, i.e. the average number of interactions in triggered events, is
\begin{equation}
\label{eq:pile}
  p = \frac{\mu}{1-f_0}
\end{equation} 
and the rate of microbias triggered events at a luminosity $\mathcal{L}$ is 
\begin{equation}
\label{eq:sig}
\frac{dN_{\text{microbias}}}{dt} = \sigma 
\times \frac{\mathcal{L}}{p}
\end{equation}
where $\sigma$ is the cross-section evaluated to be 65~mb with 10\% uncertainty, from the first luminosity
studies.
Starting from the measured values of the number of firing pads per triggered event (equation~\ref{eq:rpt} for microbias triggers), the hit 
rates for each sample are extrapolated to the nominal
luminosity of $2 \times 10^{32}$ Hz/cm$^2$, obtaining the normalized rates
\begin{equation}
\label{eq:norm}
R= r_{\text{microbias}} \times \frac{dN_{\text{microbias}}}{dt} \times 
\frac{\mathcal{L_{\mathrm{nominal}}}}{\mathcal{L}}
\end{equation}
The result is shown in Fig.~\ref{fig:rate}.
\begin{figure}[h]
  \centering
  \mbox{\includegraphics[width=.9\textwidth]{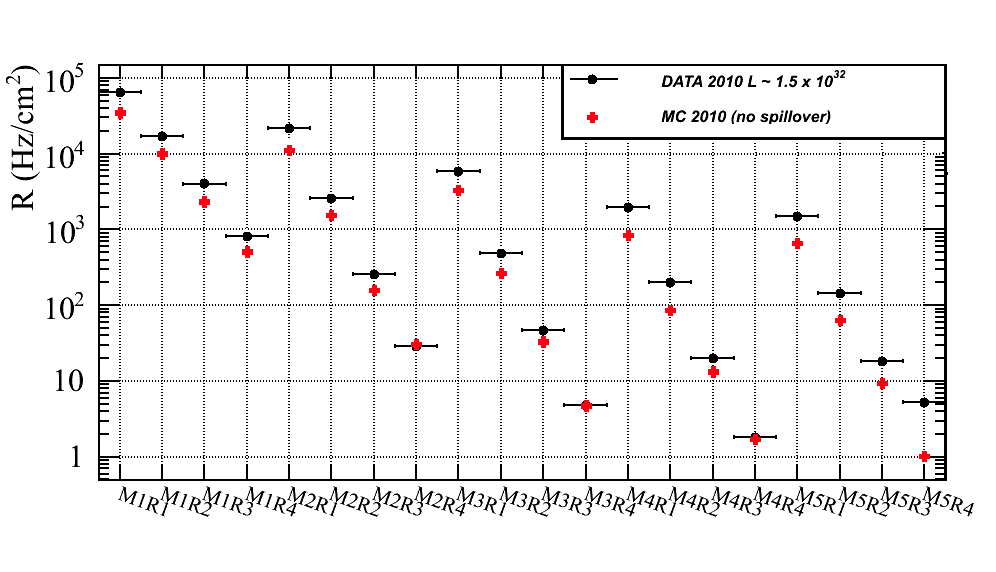}}
\vspace*{-4truemm}
  \caption{Comparison of the normalized rates $R$ (\ref{eq:norm}) seen in 
the 20 muon detector regions for a high luminosity run, 
with the MC data produced in the 2010 configuration at $\sqrt{s}$ = 7 TeV. 
The rates refer to the in-time hits (within the trigger 25 ns gate).}
\vspace*{5truemm}  
\label{fig:ratevsmc}
\end{figure}

A satisfactory scaling of the rates
is verified across five order of magnitudes in
luminosity, indicating that contributions not due to the triggered collisions are small with the only important exception of the outer regions of the last station M5 
where a significant effect from
back scattering is expected due to the limited shielding 
behind the detector.
The contribution of back scattering depends on the beam conditions 
($\mu$ and bunch spacing) and will be further discussed later in this section.

In Fig.~\ref{fig:ratevsmc} the values of the normalized rates $R$, measured in each region for a high luminosity run are compared with
the Monte Carlo data sample produced in the 2010 configuration at $\sqrt{s}$ = 7 TeV. It must be however taken into account that spillover was not simulated in this sample.
The Monte Carlo reproduces the rates of the large outer
regions R4 for M2, M3, M4 stations while a lower rate predicted for the outer region of M5 can be attributed to the missing spillover simulation (see later for a further discussion).
On the contrary  the Monte Carlo rates for M1 and all the inner regions are lower by up to a factor 2. This indicates that the track multiplicity predicted by PYTHIA~\cite{pythia}, and the contribution of low energy particles from showers around the beam pipe were underestimated in the simulation. 
The highest rates measured are  however within the safety factor assumed in the detector design phase.

\begin{figure} [t]
  \centering
  \mbox{\includegraphics[width=.75\textwidth]{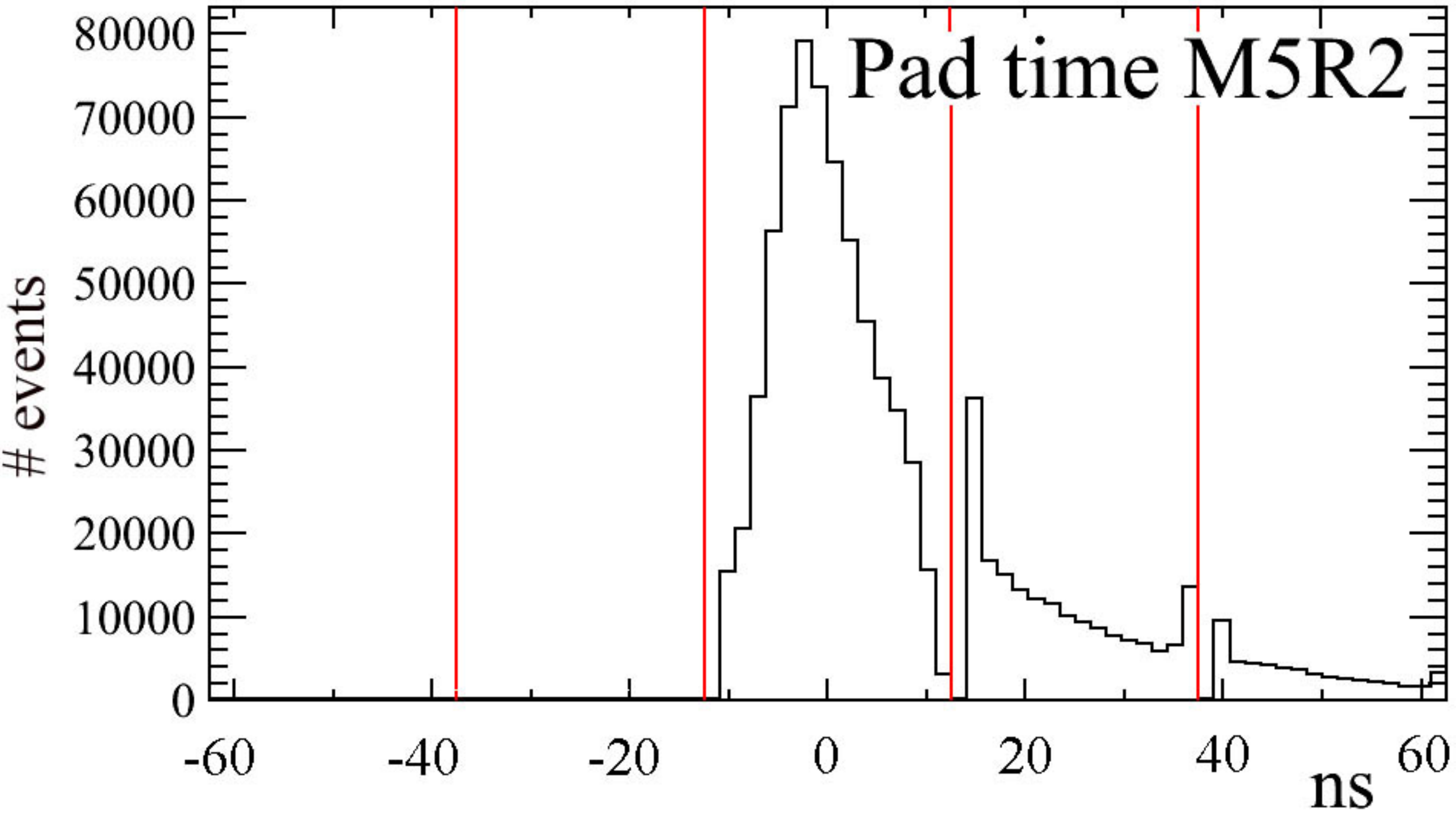}}
  \caption{ Time distribution of the M5R2 hits for  L0MB events acquired 
in TAE mode. The red vertical lines separate the  consecutive 25 ns gates 
assigned with progressive BXID numbers. The structures at the gate 
boundaries are due to a known feature of the TDC giving an incorrect fine 
time measurement at the gate edges. 
  }
\vspace*{8truemm}
  \label{fig:spillover}
\end{figure}

The spillover contribution was 
estimated from
TAE data by using the bunch crossing identification number (BXID) and the fine time information of the 4-bit TDC in the ODE boards. 
The L0MB events were used, since the microbias triggers were not available in the TAE mode.
A typical time distribution showing spillover is reported in Fig.~\ref{fig:spillover}.

\begin{figure}[t]
\vskip -5truemm
 \centering
  \mbox{\includegraphics[width=.9\textwidth]{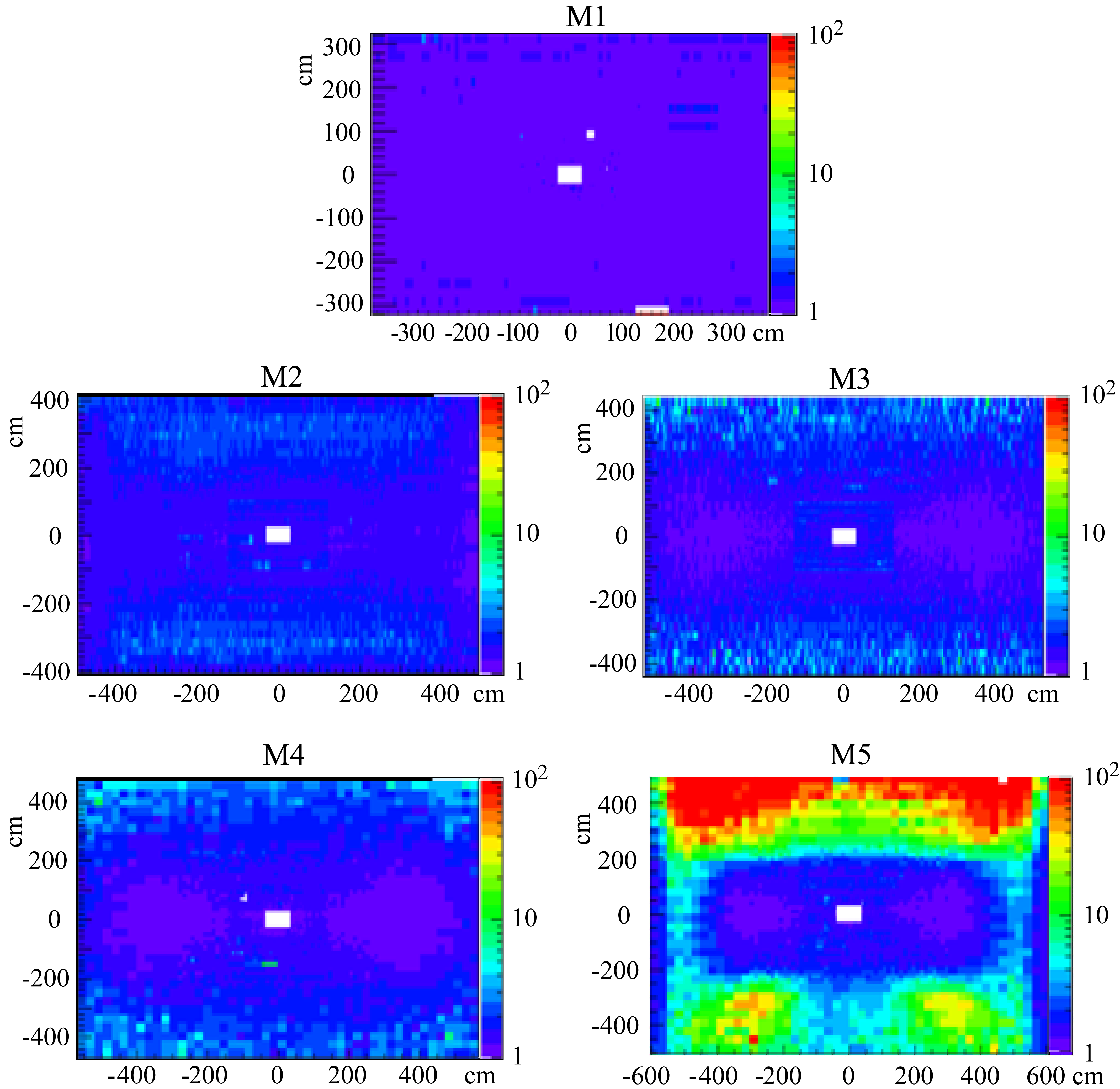}}
  \caption{Space distribution of the ratio between the hit rates measured in 125 ns
    and 25 ns. The contribution of late hits from backscattering is
    evident in the outer regions, notably for station M5.}   
  \label{fig:ratelate}
\vspace*{1truemm}
\end{figure}

The space distribution of any-time signals (125 ns gate) normalized to the in-time (within the standard 25 ns gate) signal distribution,  
is shown 
in Fig.~\ref{fig:ratelate}. 
The plots show excess of late 
hits, clearly not related to the amount of the in-time 
ones, notably in the up and down edges of the more downstream stations.  
The effect is impressive in the last station M5 where the too small iron wall behind the 
detector (see Fig.~\ref{fig:muondet}) provides insufficient shielding from backscattered particles. 
It is worth to note that the effect of late hits 
is enhanced in regions
where the rate of in-time hits is small and is suppressed where the in-time particle flux is large. This effect is visible for instance in the left and right edges of M5, not fully shielded from in-time particles by calorimeters and upstream muon stations.

\begin{figure}
  \centering
  \mbox{\includegraphics[width=0.9\textwidth]{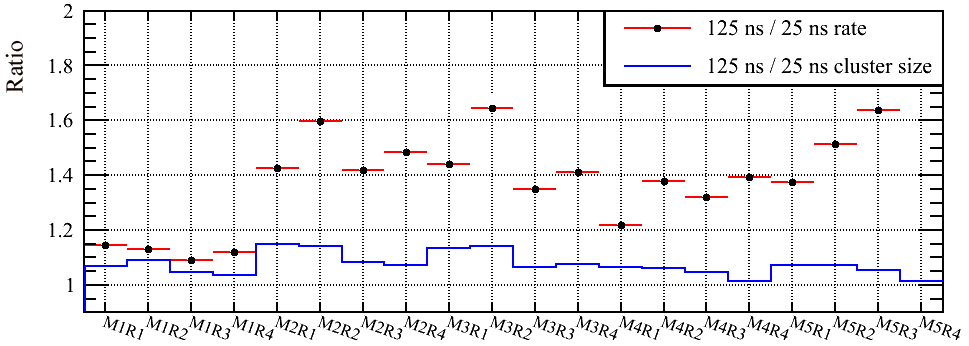}}
  \caption{
   Dots in red: ratio of the hit rate in 125 ns time window to the hit rate in 25 ns; the out of scale value for M5R4 is about 10.
   Continuous line in blue: ratio of total cluster
  size to in-time cluster size for track hits. 
    }
  \label{fig:rateTAE}
\end{figure}

In the M5R4 region the total rate of hits in TAE events 
increases by a factor 10 with respect to the in-time hit rate (from 1.5 to 15 Hz/cm$^2$). The increase for the other regions can be seen in Fig.~\ref{fig:rateTAE}.
The large effect measured for the downstream stations 
is essentially  due to late backscattered particles. 
The smaller increase seen for station M1, instead, is mostly due to late cross-talk signals from in-time 
particles\footnote{This will be clarified in section~\ref{sec:clsize} where the time behaviour of cross-talk is studied. The effect of rate increase 
due to late cross-talk hits meaured for clusters associated to a track is reported in Fig.~\ref{fig:rateTAE}.}. 
The overall spillover effect is somewhat large in terms of relative rate increase, but not worrying in terms of absolute
 occupancy level. It must be added that spillover hits, being mostly seen in the 25~ns following the
bunch crossing, do not affect significantly the measured rates when beams are operated with 50~ns bunch crossing spacing.


\begin{figure}[h]
  \centering
 \includegraphics[width=0.9\textwidth]{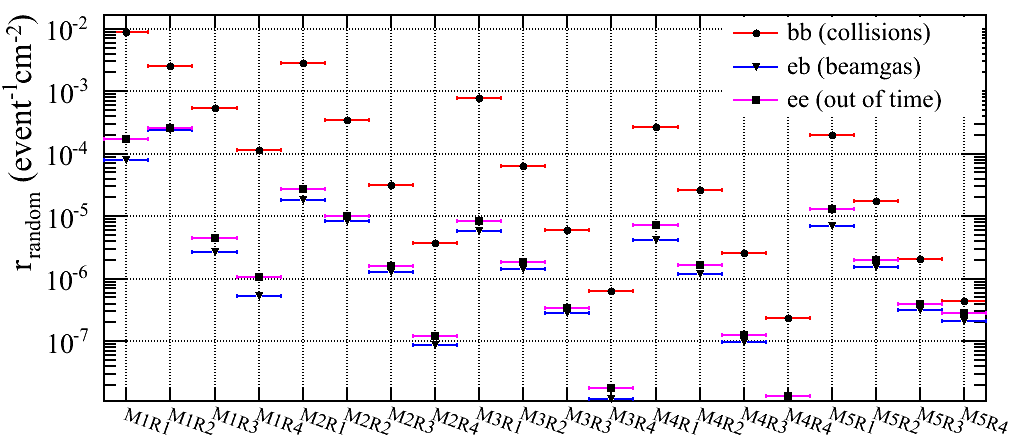}
\vspace*{2truemm} 
 \caption{Comparison of the average number of hits per cm$^{2}$ 
(equation~\ref{eq:rpt}) measured in a colliding beams run for 
random triggers in coincidence with beam-beam collisions (bb), 
non-colliding bunches (eb) and no beams (ee) -
see text for details. }   
\vspace*{5truemm} 
  \label{fig:rateOutbb}
\end{figure}

Finally, a check for possible sources of particles outside collision events 
was performed comparing the rates measured in events randomly triggered by 
the machine 40~MHz clock, in time or out of time with passing beams.
In Fig.~\ref{fig:rateOutbb} the number of hits per unit surface 
(equation~\ref{eq:rpt})  measured in a run with 150~ns spaced bunch 
trains, is reported.
Events triggered in coincidence with beam-beam collisions (bb) are 
compared with empty-beam (eb) and empty-empty (ee) events. The hit 
rate seen  in empty-beam events, triggered in coincidence with the 
passage of non-colliding bunches, is due to beam-gas interactions and 
receives 
little contribution from beam-beam collisions that are at least 150~ns 
away. The hit rate of empty-empty events, triggered out of time with 
passing beams, is instead dominated by the delayed hits from collisions  
occurring 25 or 50~ns before the triggered clock.     
Except for the low-rate external regions of M5, rates outside collisions 
are less than 10\% of the collision rates.
The hit rate measured far from LHC bunch trains is compatible with 
the residual detector noise.
\vskip 4truemm

\section{Cluster Size}
\label{sec:clsize}

The average cluster size of muon track hits is an essential 
parameter of the
detector response, since it monitors the correct 
operation conditions of the chambers
and affects the muon trigger performance. 
In section~\ref{sec:datasample} it has been described how the firing logical pads are clusterized and the track hits defined.
The cluster size can be measured in terms of the average 
number of firing adjacent 
pads.
A cross-talk between adjacent pads due to charge signal induction and to some capacitive 
coupling is expected.
The cluster size values for standard operating conditions 
were in the past measured on test benches, for particles impinging perpendicularly to the chamber plane, and used to feed Monte Carlo simulations.

The average cluster size in the $y$ coordinate is near to 1, while in $x$, due to the finer segmentation of chambers in the bending coordinate, is significantly larger than 1. The cluster size along $x$, and its dependence on the detector geometry of each region, was studied.
 
\begin{figure}[h]
  \centering
  \mbox{\includegraphics[width=.75\textwidth]{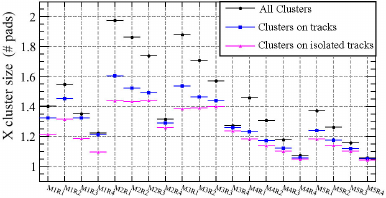}}
  \caption{Average cluster size along $x$ in a low luminosity run for each detector region. 
Events are triggered by microbias or L0MB triggers. 
Cluster
selections are described in the text. }
\vspace*{5truemm}  
\label{fig:xclsizes}
\end{figure}

It is worth to note that the average cluster size of muon track hits is smaller than the one for all the hits in the detector, as can be seen in Fig.~\ref{fig:xclsizes}. This effect is large in regions where most of the radiation is due to low energy particles.
If clusters associated to a track
are also required to be isolated\footnote{ The isolation condition requires non-adjacent 
firing pads
in the non-bending direction ($y$ cluster size =1) and no 
other firing channels
in the same station within 7 logical pads in $x$ and 2 in $y$.}, 
the cluster size is further reduced demonstrating an effect of hit coalescence. This effect is run dependent, being correlated with 
event pile-up. 
The isolation cut reduces the effect of pile-up though not fully suppressing it~\cite{bib:performance-note}. For this reason a run 
with a small $\mu$ value must be used to make a comparison with the cluster size measured on test benches.  

\begin{figure}[t]
\vskip -4truemm
  \centering
  \mbox{\includegraphics[width=.7\textwidth]{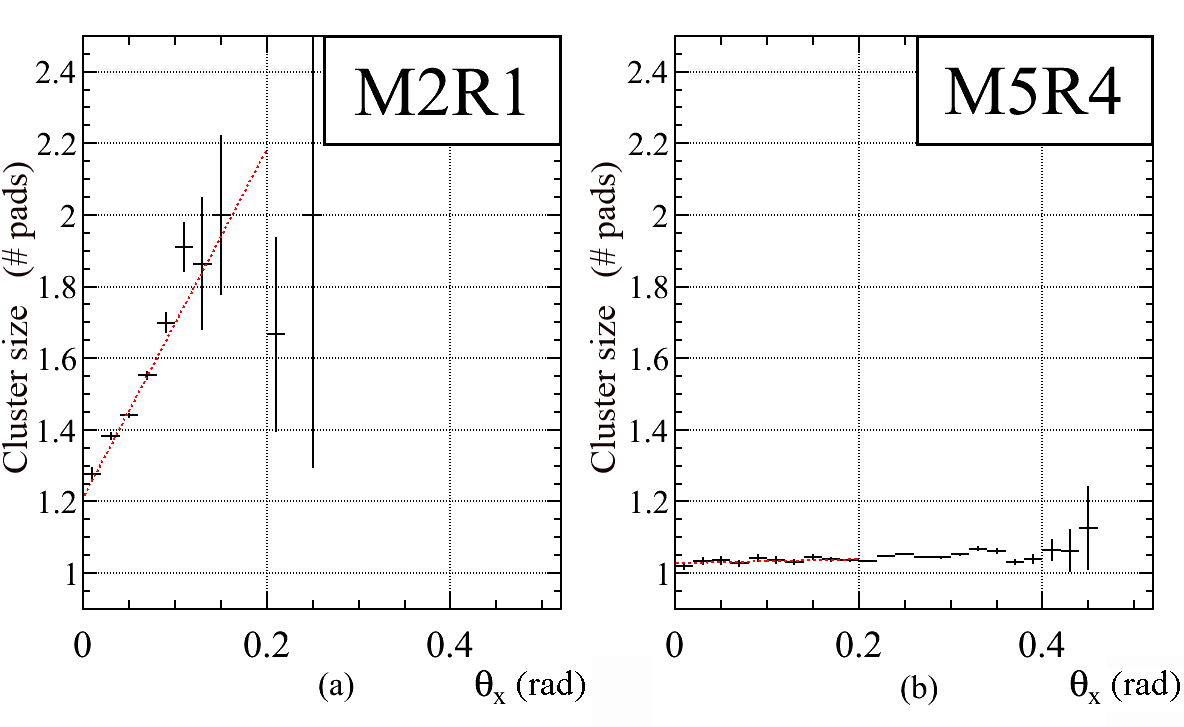}}  
\vspace*{-2truemm}
  \caption{Average x cluster size for isolated track hits as a
    function of the track angle (in rad) for M2R1 (smallest logical pad region) and M5R4 (largest logical pad region). 
    A linear fit
is used to evaluate the cluster size for perpendicularly  impinging tracks.}
  \label{fig:CLS-angle}
\end{figure}

\begin{figure} [h]
\vskip -6truemm
\centering
  \mbox{\includegraphics[width=.85\textwidth]{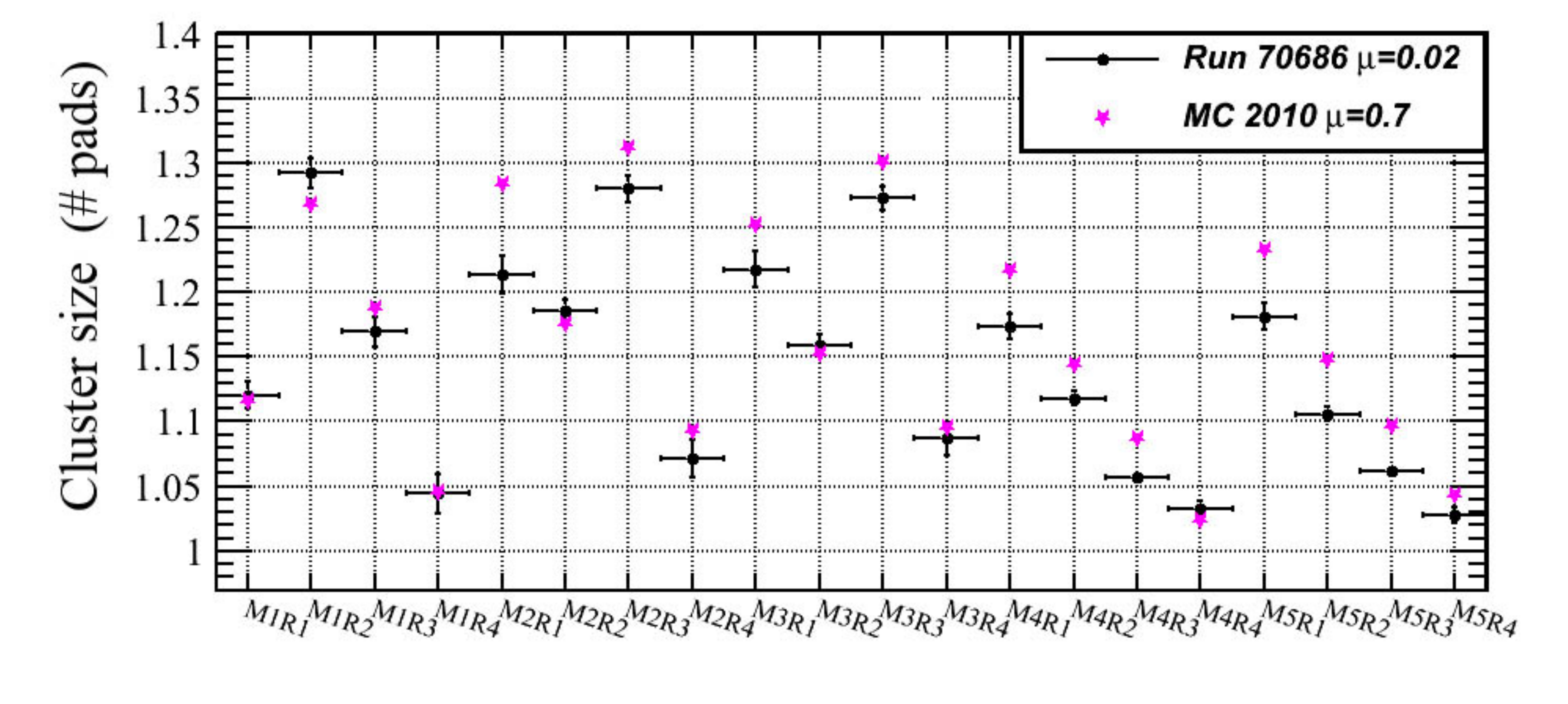}}
\vspace*{-5truemm}
  \caption{Average $x$ cluster size at 0 angle in experimental data and simulation (MC 2010). To suppress the effect of pile-up, only isolated clusters and low-luminosity data are used.}
  \label{fig:xclsizeCmp}
\end{figure}

The purely geometrical effect due to the muon trajectory inside chambers
was measured by plotting the average $x$ cluster size 
as a  function
of the angle between the track projection on the bending 
plane and the perpendicular to the
chamber plane. A correction for this effect was performed by extrapolating 
to 0 angle, as shown in
Fig.~\ref{fig:CLS-angle}.
The resulting average cluster sizes are reported in Fig.~\ref{fig:xclsizeCmp} 
and compared with the  values obtained in the simulation. The relative 
behaviour in the different regions is quite well reproduced and the cluster 
size in the real data is  in several cases smaller, indicating a better 
tuning of the chamber working conditions since the time of the test benches.
 The cluster sizes never exceed 1.35, a value well within the L0 trigger 
requirements~\cite{bib:l0mudesign}.

\begin{figure}[b]
\vskip -30truemm   
  \centering
  \mbox{\includegraphics[width=.75\textwidth]{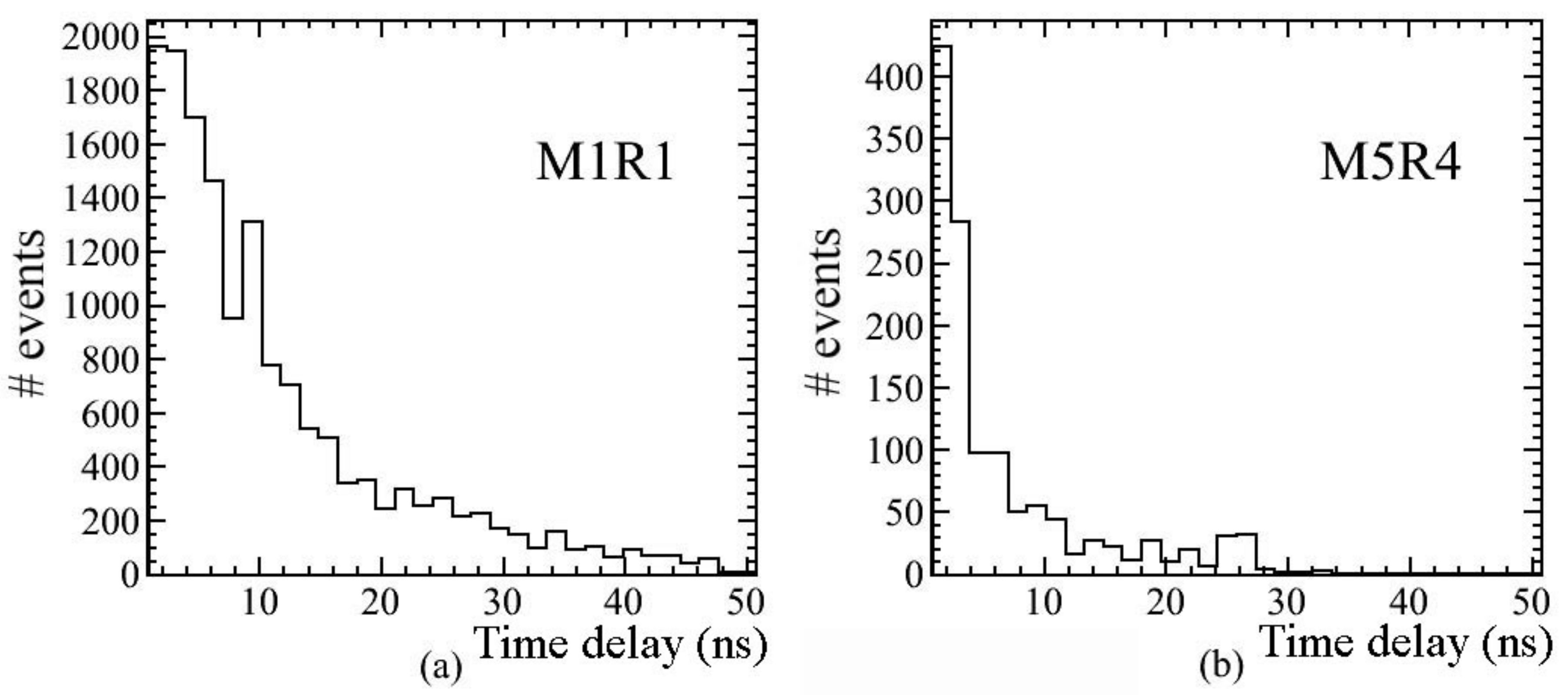}}
\caption{Time delay distribution of pads in track clusters with 
  respect to the
    first pad in time: (a) for the small cathod pad region M1R1; (b) for the large anode pad region M5R4.  }  
\vspace*{0truemm}
 \label{fig:xcltiming}  
\end{figure}

The time behaviour of cross-talk hits was studied using TAE data.
While the time distribution of the first pad in time in a cluster is almost fully contained in a 25~ns time window, so ensuring a high trigger efficiency, the other pads in the cluster can arrive significantly later
as it is shown in Fig.~\ref{fig:xcltiming}.  
As a consequence, the cluster size (as the hit rate) measured  in a 125~ns
time window is larger with respect to the one measured in the 25~ns window. The importance of the effect is region dependent and is quantified in Fig.~\ref{fig:rateTAE} for clusters associated to muon tracks.
The late cross-talk signals are not normally
acquired and thus do not affect the trigger  performance, though
contributing to background for future beam operations with 25 ns bunch
spacing.
\clearpage

\rule{5cm}{0pt}
\clearpage

%% file: timing.tex
\section{Timing}
\label{sec:timing}

The L0MU trigger requires muon hits to be
recorded in each of the five stations within the 25~ns LHC gate associated to
a beam-beam crossing. This timing constraint is the most stringent
requirement for achieving  the design 95\% muon detection
efficiency. In fact the tails in the chamber time response
are expected to be one of the main sources of detector inefficiency.
To reach the goal, the detectors were conceived to have a
time resolution better than 4 ns at their nominal settings, while the 122k readout channels have
to be time-aligned at the 1~ns r.m.s. level.
 
The time alignment of the detector has been achieved in several
steps. Test signals produced by a custom pulser system~\cite{bib:bocci}
 were used for
a first timing. Cosmic data collected in 2009 allowed to refine
the channel equalization using physical signals. These two steps are described in
detail in \cite{bib:cosmicpaper} and \cite{bib:cosmictimenote}. 
After the calibration with cosmic data, a
satisfactory time resolution was already 
reached for all regions except the most
inner ones where the statistics was a limit.
The precise timing of the first beam particles allowed to quickly 
intercalibrate the channels of the highly illuminated inner regions. 
For the other regions, only a few channels exhibiting an
anomalous shift in the time response, due to some hardware
interventions, were identified and fixed. 

The detector efficiency could then be optimized by
a fine tuning of the channels time offset 
with respect to the 40 MHz LHCb clock. 
The single channel time distribution
exhibits an asymmetric shape, with a longer tail for late times.
This is due to the intrinsic chamber response 
(dependent on drift time and time walk with pulse height), 
to the effect of delayed cross-talk hits and 
to the effect of the longer path of low momentum tracks. 
This implies that, in order to minimize the fraction of signals
falling outside the 25 ns gate, the average time should not be
centered on the middle of the DAQ gate, but slightly before.
The optimal offset is region dependent as is the shape of the time
spectrum. 
Moreover, small shifts among regions were already introduced by
the optimization of the HV and threshold settings 

The offset optimization 
was then independently performed for
each detector region. Special runs were acquired with L0MB triggered events in TAE
mode, varying the global time offset in steps of 1 ns. For each
data sample, standalone muon tracks were reconstructed requiring hits
in all five stations. 
The optimal offset was chosen by maximizing the
timing efficiency, defined as the
probability that at least one of the track hits in a given station is found
within the central \hbox{25~ns gate~\cite{bib:performance-note}.} 

\vspace*{5truemm}
\subsection{Time Performance}\label{sec:timingresults}

The detector time performance for each detector region was estimated~\cite{bib:performance-note} by analysing two TAE event samples:
\begin{itemize}
\item The first sample was acquired at the beginning of the physics data taking (before applying the final intercalibration for 
region M1R1) with events triggered by L0MB;
\item  The second sample was acquired at the very end of physics data taking. At that time the L0MB triggers were downscaled by a 
factor 100 and the need of an adequate statistics required the use of the
physics L0 triggers based on calorimeters (electron, photon, hadron
triggers).
\end{itemize}
TAE events were fully reconstructed so that 
M-tracks could be  required to have a good
matching with a T-track having a momentum larger than 8 GeV/c to ensure muons
to reach the M5 station. 
Ttracks were required to cross the muon detector at a safety distance from the inner and outer edges to avoid border effects. 
Clusters inside the few chambers with pathologic behaviour mentioned in section~\ref{sec:functioning}, were not considered for the timing efficiency calculation.  

\begin{figure} [h]
  \centering
  \mbox{\includegraphics[width=.75\textwidth]{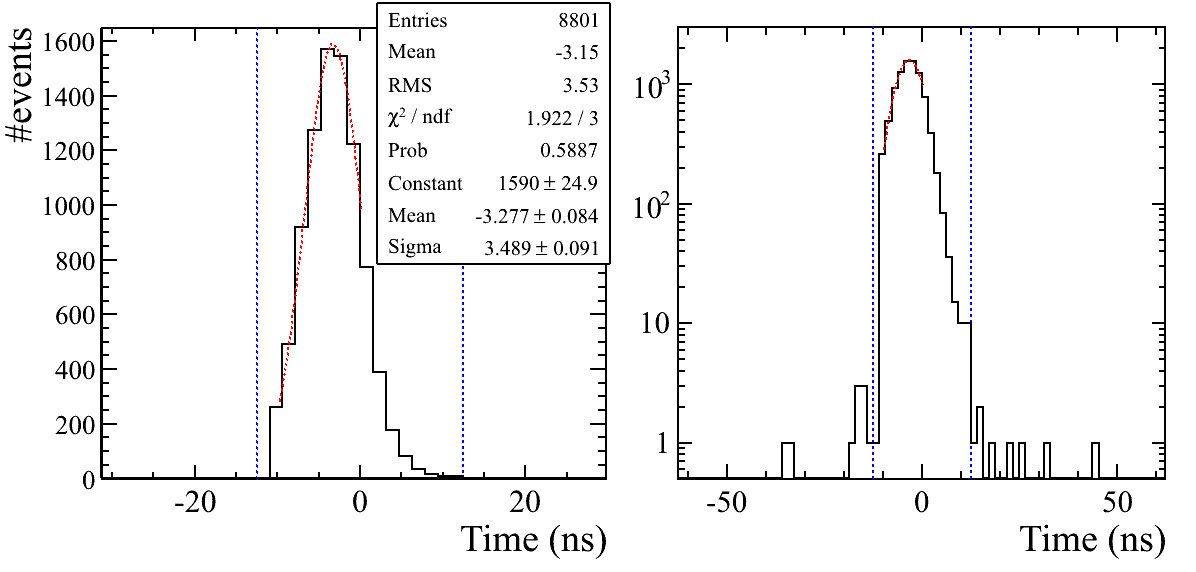}}
  \caption{ Time distribution of the most time centered hit, measured in M5R2 for L0MB events acquired in TAE mode, reported in linear (left) and logarithmic (right) scale.   
The vertical lines at $\pm$ 12.5~ns delimit the ``efficient'' hits 
assigned with the BXID number of the trigger. The result of a gaussian fit performed around the maximum of the distribution to evaluate the core time resolution is also shown.
  }
  \label{fig:firstHITmu}
\end{figure}

The time resolution and, most importantly, the timing efficiency is evaluated from the distributions obtained with the BXID and the 
4-bit TDC mesurement
for the most time centered hit in the track clusters.
If such hit is not assigned with the correct BXID , even if the detector is firing, we have an inefficiency due to the timing.
As an example, the distribution obtained with the second TAE sample for the M5R2 region is reported in Fig.~\ref{fig:firstHITmu}.
From these distributions, the core time resolution was also evaluated from 
a gaussian fit in the interval (-6, +4)~ns around the 
maximum and its error estimated by shifting the 
fit interval by $\pm$~2ns.” 

The average timing efficiency of the five stations and each of the 
20 regions, obtained for the TAE sample acquired at the end of data 
taking, is reported in Tab.~\ref{tab:timeff}. 

\vspace*{3truemm}
\begin{table}[b]
\begin{center}
\caption{Average timing efficiency, in percent, for the five stations and for each region, obtained from the TAE sample acquired at the end of data taking. Statistical errors have been evaluated with an approximated binomial 68\% confidence interval.   } 
\vspace*{4truemm}
\renewcommand{\arraystretch}{1.5}
\begin{tabular}{|c|c|c|c|c|c|} \hline
Station& ~~~~~~~~R1~~~~~~~~ & ~~~~~~~~R2~~~~~~~~ & ~~~~~~~~R3~~~~~~~~ & ~~~~~~~~R4~~~~~~~  & Station efficiency\\ 
\hline
 M1 & $ 98.42\;{ ^{+\,0.19}_{-\,0.39}}$ & $ 99.50\;{ 
^{+\,0.05}_{-\,0.10}}$ & $ 
99.78\;{^{+\,0.03}_{-\,0.10}}$ & $ 99.77\;{ ^{+\,0.03}_{-\,0.26}}$ & 
$99.51\;{^{+\,0.04}_{-\,0.06}}$ \\
\hline
M2 & $ 99.56\;{ ^{+\,0.06}_{-\,0.35}}$ & $ 99.72\;{ ^{+\,0.04}_{-\,0.09}}$ 
& $ 99.91\;{ 
^{+\,0.01}_{-\,0.07}}$ & $ 99.89\;{ ^{+\,0.02}_{-\,0.21}}$ & $99.80 \;{
^{+\,0.02}_{-\,0.04}}$ \\ \hline
M3 & $ 99.48\;{ ^{+\,0.07}_{-\,0.41}}$ & $ 99.72\;{ ^{+\,0.04}_{-\,0.09}}$ 
& $ 99.91\;{ 
^{+\,0.01}_{-\,0.06}}$ & $100.00\;{ ^{+\,0.00}_{-\,0.17}}$ & $99.83\;{ 
^{+\,0.02}_{-\,0.04}}$ \\
\hline
M4 & $ 99.80\;{ ^{+\,0.04}_{-\,0.40}}$ & $ 99.96\;{ ^{+\,0.01}_{-\,0.06}}$ 
& $ 
99.88\;{ 
^{+\,0.02}_{-\,0.07}}$ & $ 99.95\;{ ^{+\,0.02}_{-\,0.18}}$ & $99.92\;{ 
^{+\,0.01}_{-\,0.03}}$ \\
\hline
M5 & $ 99.67\;{ ^{+\,0.05}_{-\,0.47}}$ & $ 99.67\;{ ^{+\,0.04}_{-\,0.10}}$ 
& $ 99.84\;{ 
^{+\,0.02}_{-\,0.07}}$ & $ 99.86\;{ ^{+\,0.02}_{-\,0.20}}$ & $99.77\;{ 
^{+\,0.03}_{-\,0.05}}$ \\
 \hline
\end{tabular}
\vspace*{4truemm}
 \label{tab:timeff}
\end{center}
\end{table}

It can be noted that efficiency values in M1 are lower than in the other stations. This is expected considering that M1 is equipped with bi-gap instead of quadri-gap MWPCs and the GEM detectors in the inner region M1R1 have an intrinsic poorer time resolution. Also in the high granularity regions M2/3 R1/2 larger inefficiencies can be expected due to the presence of double readout chambers where 
the logical ``and" of two signals is required. 
Nevertheless the probability for a muon track reconstructed in TAE events 
to have hits ``in time" in all the 5 stations (overall timing efficiency)  
is measured to be 98.83~$\pm$~0.09\%, a value well beyond requirements.
The measured values exhibit a sensitivity to the quality 
cuts suggesting that the efficiency could be systematically
underestimated, by a few per mill, in the regions most affected 
by combinatorial background, namely region R1 and stations M1 and M5. 
On the other hand, the statistics of the available data samples 
does not allow to tighten the cuts, further improving the 
muon track sample purity. This is particularly evident for the crowded regions R1 where the statistics of tracks in TAE events, essentially containing low momentum muons from $\pi$ decays, is limited. Moreover a  further reduction of the sample is due to a fiducial volume cut removing tracks  near the beam pipe. This cut is needed to reduce the probability of reconstructing tracks with accidental background hits that are characterized by a much wider time distribution.

\begin{figure} [h]
  \centering
\mbox{\includegraphics[width=1.0\textwidth]{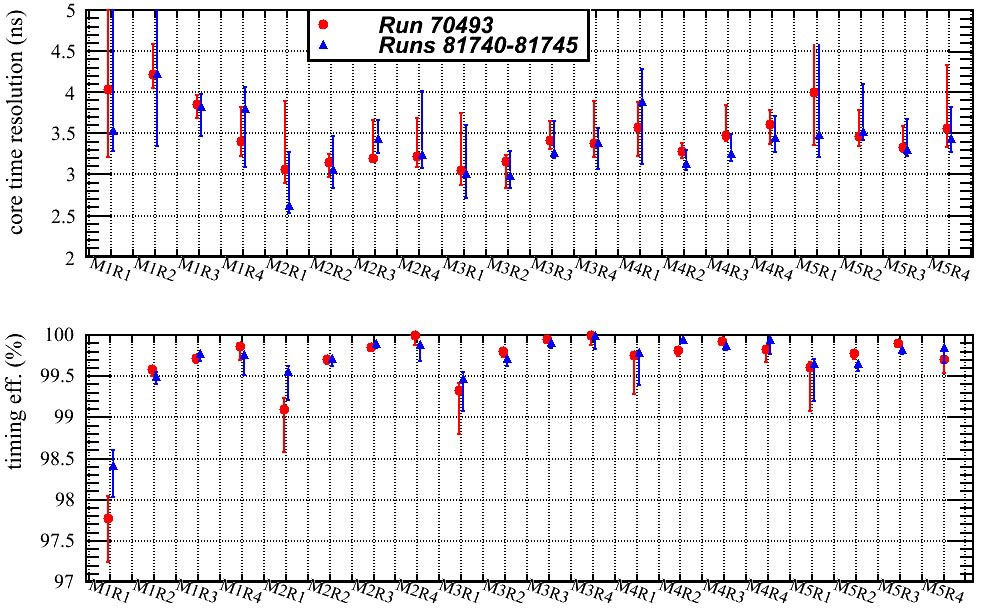}}
  \caption{
Core time resolution (upper plot) and timing efficiency 
(lower plot) measured for each region  
  in the two
    TAE samples acquired before and after the
    bulk of 2010 LHCb physics data.  }
\vspace*{5truemm}  
\label{fig:tTAEcmp}
\end{figure}

The core resolutions and the timing efficiencies obtained for the two TAE samples are compared in Fig.~\ref{fig:tTAEcmp}.
Despite the different triggers used in the two samples, resulting in a different momentum spectrum 
and space distribution of the tracks, the results were found to be in very good
agreement, the hint of difference for region M1R1 can be attributed to a more accurate time calibration used for the second sample.
This result demonstrates the excellent stability of the muon
system along the 2010 run. 

\subsection{Stability of the time response}

\begin{figure}[htbp]
  \centering
  \mbox{\includegraphics[width=0.65\textwidth]{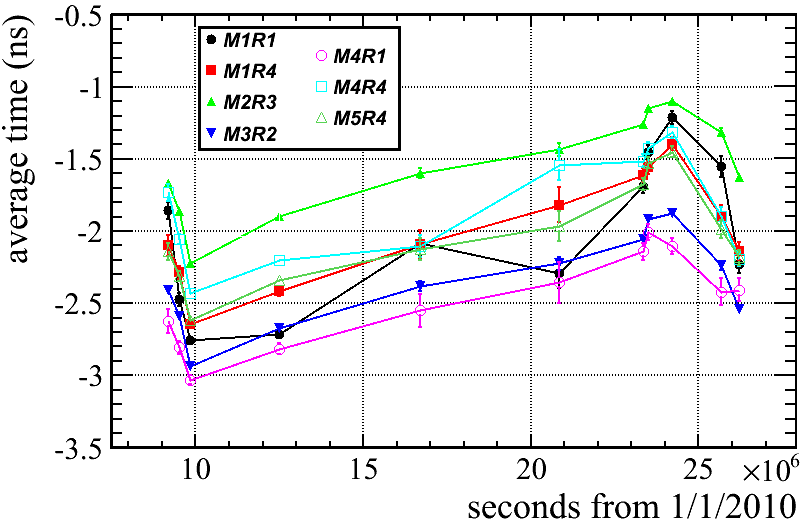}}
  \caption{Average time of the most time-centered track hit
for different stations and regions, measured in the sample runs
as a function of their data-taking time.
The values for optimal efficiency depend on the detector region and have
  been fixed at the start of data taking. Variations along the 7 months of
  operation do not exceed the  $\pm$~1~ns range.
}
\vspace*{5truemm}
  \label{fig:tStab1}
\end{figure}

The stability of the absolute time scale in the long term is
expected to be limited by two effects:
\begin{itemize}
\item the LHCb clock drifts with temperature; variations were
  compensated manually during the run in order to be stable within
  $\pm$~0.5~ns;
\item the variations of temperature and atmospheric pressure affect the chamber gain and variations of signal pulse height
produce time walk effects. The largest effect is expected from
pressure variations, and is estimated to be equivalent to a $\sim
\pm$~20~V change in HV~\cite{bib:pinciusnim}, corresponding to  $\sim \pm$~0.4~ns.
\end{itemize}
The average time of track hits, measured
in the sample runs acquired across
the 2010, is shown 
in Fig.~\ref{fig:tStab1} for various detector regions as a function of the data taking time. 
The behaviour is consistent with the mentioned effects.  
Variations are at the level of $\pm$~1~ns and are clearly correlated among regions. Residual uncorrelated variations are compatible with zero.
There is no evidence for a dependence of the time drift on the detector
illumination, that could be a hint for an ageing effect on the chamber gain. 

%% file: align.tex
\section{Spatial alignment}
\label{sec:align}

The spatial alignment of the muon detector must guarantee the design performance of trigger and  offline muon identification. 
As explained in section~\ref{sec:muondet}, the L0MU trigger requires hits in all the 5 stations aligned on a muon track-segment having a $p_{T}$  above a given threshold.
Offline muon identification is more flexible in the reconstruction of the muon track-segment but requires the matching with a track reconstructed in the tracking system  through the whole spectrometer.

The alignment accuracy needed in the system is driven by the 
trigger requirements in the stations M1, M2 and M3. For these three stations 
the FOI of the hit search window in the non-bending vertical coordinate, 
is defined by 1 pad only. As a consequence, a relative $y$ misalignment 
between the stations would directly contribute to trigger inefficiency ($\sim$2\% for 1mm misalignment).
In the bending coordinate $x$, the FOIs are composed of several  pads and the main effect of a misalignment is a bias in the $p_{T}$ calculation, 
at the percent level for 1mm misalignment. However the bias can be removed if the trigger algorithms are suitably corrected using  the true $x$ positions.
The alignment of stations M4 and M5 is less important because in their case the $y$ FOI is as large as 3 pads and their hits are not used to calculate $p_{T}$.
 The detector mechanics was designed with the aim of reaching a precision of the order of 1~mm in $x$ and $y$ 
 directions. The alignment requirements along $z$ are much less demanding due to the forward geometry of the experiment.

\begin{figure}[!ht]
\centering
\includegraphics[width=.85\textwidth]{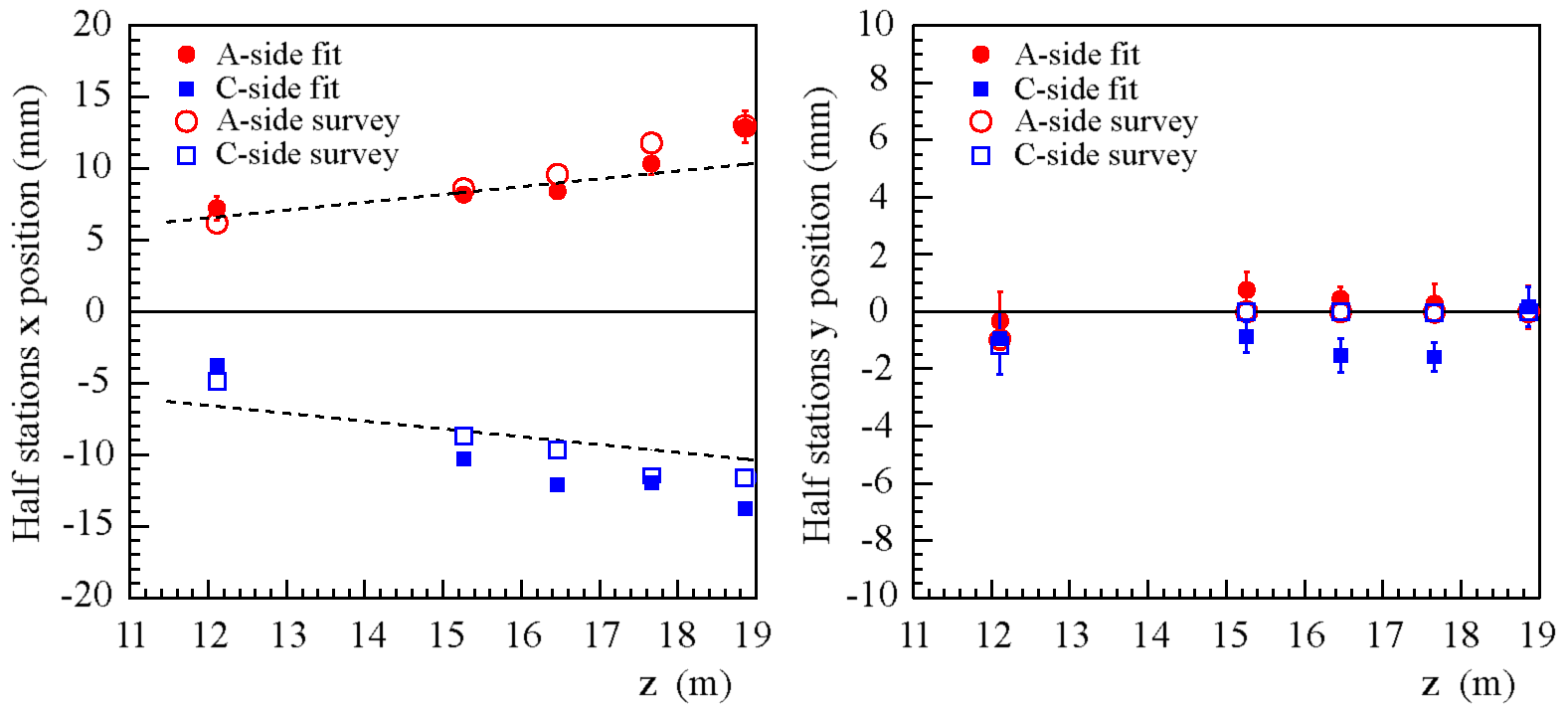}
	\caption{
Alignments of the ten muon half stations
for the 2010 run. 
The average value $x$ of the inner edges (left) and the 
median $y$ (right), are shown 
as a function of the station $z$ position.
The empty dots represent the survey 
measurements whose errors are negligible. The full dots are the positions found by the software global alignment described in section~\ref{sec:alignA}; the error bars correspond to the statistical and systematic uncertainties, summed in quadrature. The dashed lines in the left plot represent the ideal alignment.}
	\label{fig:Align2010}
\end{figure}

During the installation, the supporting walls were kept in the open position and the muon chambers mounted with an accuracy of 
$\sim$1~mm along $x$ and $y$ coordinates,
 centred on their nominal positions, relative to reference targets placed on top of each half station. 
The measured rotations were zero within the precision of 1~mrad.  
After  chamber installation, the half stations were closed around the beam pipe leaving a small safety distance between the A and C side;
the two half stations being ideally positioned
left-right symmetric and projective to the interaction point.  
The closed stations were then surveyed with respect to the LHCb cavern using  four reference targets on each side, and the values stored in the geometry database used by the offline reconstruction program to define the absolute hit coordinates. The values measured by the survey for the 2010 collision run
 are  reported in Fig.~\ref{fig:Align2010}. They show non negligible though small misalignments from the ideal position for the M1C half station. The misalignments of the other stations are negligible or unimportant.

\subsection{Space alignment measured with muon tracks}
\label{sec:alignA}

A study of the muon system alignment was performed using muon tracks with the first purpose of checking the mechanical positioning and the survey measurements. In the same time a tool was prepared for alignment monitoring after each intervention requiring the opening of the stations, and eventually for correcting the geometry data base.  

In this analysis M-tracks
are defined by at least four clusters
in four different stations that are compatible with a 
straight line.
Candidates with too large clusters (number of pads$>$6) are 
eliminated to avoid mis-reconstruction problems. Moreover 
stations with more than 300 hits are excluded. 
The M-tracks are required to match a good quality  T-track having 
a momentum 
$p>$6 GeV/c. The matching condition requires a good 
$\chi^{2}$ between the parameters of M-track and T-track 
extrapolated to the M2 position. Matching segments are then 
merged together in a unique track that is required to 
have a good $\chi^{2}$. 

The alignment is performed following the standard procedure used in 
LHCb~\cite{bib:Kalman}. It makes
use of the \emph{Kalman fit iterative method}, that performs a 
minimization of the total $\chi^{2}$ of an ensemble of 
tracks while adjusting the detector positions. 
 Only the positions of the muon half stations were allowed to vary 
 since the tracking detectors were previously  
 independently aligned using the same procedure. 
The iterative process starts assuming the muon detector 
in the positions measured by the survey and stops  when the convergence is reached (total $\chi^{2}$ doesn't improve significantly) usually after 4-5 iterations.  
In Tab.~\ref{tab:GlobalFitTxTy} are reported the misalignments
relative to the survey, measured on a sample of $\sim$~7000 
tracks, with an analysis performed fitting only the
translational degrees of 
freedom along $x$ and $y$ of each half station.
The absolute positions in the LHCb reference system are shown in Fig.~\ref{fig:Align2010} together with the survey measurements.
 
\begin{table} [t]
\begin{center} 
\caption{Misalignments of muon half stations M1--M5, 
relative to the survey measurements,
calculated in the 
LHCb reference system with the Kalman fit iterative 
method. 
The quoted uncertainties are the fit errors (first) and the systematic 
uncertainties (second) on the relative positions determined repeating 
the analysis with different track selections. } 
\vspace*{4truemm}
\label{tab:GlobalFitTxTy}
\begin{tabular}{|c| c c | c c |}
 \hline 
&\multicolumn{2}{c|} {{\vphantom{I$^{I^I}$}} C-side}&
\multicolumn{2}{c|}{A-side}\\
& $\Delta$x (mm) & $\Delta$y (mm) & $\Delta$x (mm) & $\Delta$y (mm)   \\
M1 & 0.92$\pm$0.14$\pm$0.15  &    
0.11$\pm$0.28$\pm$0.24 &    
1.04$\pm$0.14$\pm$0.24  &  0.68$\pm$0.27$\pm$0.20 \\ \hline 
{\vphantom{$I^{I^I}$} M2} &   --1.56$\pm$0.05$\pm$0.04 &  
--0.89$\pm$0.12$\pm$0.13 & 
--0.42$\pm$0.05$\pm$0.07 &    0.76$\pm$0.12$\pm$0.17 \\ \hline
{\vphantom{$I^{I^I}$}M3} &   --2.41$\pm$0.08$\pm$0.05 &  
--1.53$\pm$0.14$\pm$0.14 & 
--1.19$\pm$0.08$\pm$0.06 & 0.45$\pm$0.14$\pm$0.08 \\ \hline
{\vphantom{$I^{I^I}$}M4} &   --0.33$\pm$0.15$\pm$0.11  & 
--1.59$\pm$0.17$\pm$0.09 & 
--1.44$\pm$0.14$\pm$0.20 & 0.28$\pm$0.17$\pm$0.15 \\ \hline
{\vphantom{$I^{I^I}$}M5} &   --2.14$\pm$0.18$\pm$0.12 &    
0.18$\pm$0.20$\pm$0.11 & 
--0.07$\pm$0.18$\pm$0.20 & 0.16$\pm$0.20$\pm$0.15 \\ \hline
\end{tabular}
\end{center} 
\end{table}

The systematic errors have been estimated repeating the analysis with samples of tracks having different momenta, comparing subsamples of tracks hitting different detector regions, and using slightly different  alignments for the tracking system.
While the absolute positions of the stations show variations of the order of  1 mm both in $x$ and $y$, their relative positions are more stable. 
The results obtained with the Kalman fit iterative method show  significant, even though small, misalignments relative to the survey measurements (in particular for side C). The geometry data base was consequently updated.
Studies on the additional degrees of freedom like rotations around $y$ or $x$ directions give results compatible with the survey.
Due to the forward geometry of the experimental apparatus, shifts in $z$ direction are as difficult to measure as unimportant, therefore
the $z$ values measured by the survey were assumed.

%% file: efficiency.tex
\section{Detector efficiency} \label{sec:effi}

The overall performance of the muon detector is quantified by the
detection efficiency of muon tracks when the system is 
operated in the standard data taking conditions.
The inefficiency introduced by dead channels, or other hardware failures occurred in the 2010 detector operation, has been quantified in section~\ref{sec:functioning}.
Here the intrinsic efficiency of the system is evaluated after applying strict fiducial volume cuts
and eliminating the few small zones where known problems are present.
An important contribution to inefficiency is expected to come from the time resolution and the time gate of 25~ns. These effects were already studied in section~\ref{sec:timingresults} and will be further discussed at the end of section~\ref{sec:results}.
Other contributions can come from intrinsic chamber inefficiencies, or  small geometrical losses due to the chamber-wise discrete structure of the muon detector.

The used procedure is described in detail in reference~\cite{bib:performance-note} and is here summarized:
\begin{itemize}
\item  
Different data sets are used for M1 and M2--M5 stations to select appropriate samples of trigger unbiased standalone muon tracks.
To reach the needed purity of the sample, M-tracks are matched with a good quality, high momentum T-track and required to fulfill tight selection criteria (section~\ref{sec:method}).
\item   
The presence of background hits affects the results and requires different procedures to correctly evaluate the true efficiency for M2--M5 stations and for M1 station where the occupancy is much higher (section~\ref{sec:bg}). 
\item  
The efficiency for each station is estimated 
by  searching clusters around the prediction defined by the M-track reconstructed 
using only the other 4 stations.
The search of clusters around the prediction is repeated 
increasing the opening window from 1 up to
8 standard deviations, both in $x$ and $y$. The value of $\sigma$
being determined, region by region, by a gaussian fit to the central part
of the distribution of the distance between the position predicted 
by the M-track and all clusters in that region.
For M1 the prediction is defined by the T-track associated to the M-track
in order to improve its quality.
The values of $\sigma$ for the twenty regions are  reported in 
Tab.~\ref{tab:allsigma}.
\end{itemize}

\begin{table}[h]
\begin{center}
\protect\caption{Resolution along $x$ and $y$ of the distance between the 
muon track and the muon cluster in each region of the muon detector. The 
muon track is reconstructed skipping the station whose resolution 
must be evaluated. }
\vspace*{4truemm}
\protect\label{tab:allsigma}
\begin{tabular}{|c|c|c|c|c|c|c|} 
\hline
 & & \vphantom{$I^{I^{I^2}}$} M1 & M2 & M3 & M4 & M5 \\
\hline
\vphantom{$I^{I^{I^2}}_A$} R1 & $\sigma_{x}\times \sigma_{y}$ (mm$^2$)&$4 
\times 
10$ &$15 \times 30$ &$10 \times 12$ &$15 \times 16$ & $33 \times 40$
 \\
\hline
\vphantom{$I^{I^{I^2}}_A$} R2 & $\sigma_{x}\times \sigma_{y}$ (mm$^2$)& $8 
\times 
18$ &$25 \times 50$ &$15 \times 24$ &$27 \times 32$ &$50 \times 60$
\\
\hline
\vphantom{$I^{I^{I^2}}_A$} R3 & $\sigma_{x}\times \sigma_{y}$ (mm$^2$)& 
$16 
\times 40$ 
& $35 \times 70$ & $25 \times 48$& $48 \times 64$ & $100 \times 110$\\
\hline
\vphantom{$I^{I^{I^2}}_A$} R4 & $\sigma_{x}\times \sigma_{y}$ (mm$^2$)& 
$32 
\times 80$ 
& $60 \times 100$ & $40 \times 96$ & $97 \times 128$& $150 \times 180$\\
\hline
\end{tabular}
\end{center}
\end{table}

If the full procedure is correct, 
the efficiency measured as a function of the
search opening window will show a saturation behaviour permitting a reliable estimate of the detector efficiency. Particular care is required in the background 
subtraction that must be a stable and reliable 
procedure even when a search area as large 
as 16~$\sigma_x\times$16~$\sigma_y$ is considered. 
 
\vspace*{5truemm}
\subsection{Muon samples and track selection}
\label{sec:method}

Different data samples are used for
M1 and for M2--M5 stations:
\begin{itemize}
\item 
For the efficiency of M2--M5 stations,
data acquired in two fills, corresponding to an integrated luminosity of 1.2 $\rm{nb^{-1}}$, were used. 
The first level trigger L0 required a high $p_T$ hadron or lepton detected in the 
calorimeter  or in the muon system; the software trigger 
HLT required the logical ``or'' of several 
independent algorithms.
To remove the bias introduced by the trigger in the 
efficiency calculation, events where both L0 and HLT were fired 
irrespectively of the muon system information were selected.
With this data sample, the majority of the muons reaching the 
muon stations and used for the analysis,  
originate from decays 
in flight of $\pi$'s or K's.   
\item 
For the M1 efficiency measurement, kaons decaying 
at the end of the tracking system can generate a good T-track 
giving  a poor quality M1 prediction, not adequate to the  large occupancy of the station. 
To have a sample of true muons, events with a reconstructed \hbox{$J/\psi\rightarrow \mu^+ \mu^-$} were used.
This sample corresponds to almost all data acquired in 2010 
\hbox{($\sim$ 37 $\rm{pb^{-1}}$)}.  
To remove the L0 bias on the efficiency evaluation, in each
$\mu^+\mu^-$ pair from a $J/\psi$, the muon which fired the L0
trigger was not considered 
in the analysis.
Notice that the use of the $J/\psi$ sample for the analysis
of the M1 station is possible because its information is not used in HLT and  reconstruction, while it would not be possible for stations M2--M5 since their information is used to identify and reconstruct both muons.
\end{itemize}
A tight selection is required 
to reach the purity of the M-track sample needed for 
 a precise efficiency measurement and different conditions are required for the different stations. 

For every station, when its efficiency is being evaluated, the M-track is validated by requiring the matching with a T-track in the other four stations used for the fit. The matching requires that the distance from the T-track extrapolated to the station and the clusters associated to the 
M-track is within one standard deviation in both $x$ and $y$ projections. Such standard deviations are estimated 
by a  gaussian fit to the central peak of the distance distributions.

A  momentum cut of 12~GeV/c (15~GeV/c) is applied to
 the T-track when M2 (M3, M4, M5) station is analysed. 
 If more than one T-track matches the M-track 
candidate and at least one of them 
has a momentum  below the cut, the candidate is rejected.
When evaluating the efficiency of M2 and M3, where the occupancy is relatively high and the fired hits are identified by crossing vertical and horizontal strips, further 
cuts on the local hit multiplicity are applied to 
avoid ghost combinations.
Moreover, when analyzing the efficiency of M3 where the prediction 
resolution is poor due to the lower granularity of M4, 
a cluster size of 1 is required on M2 station.
For the analysis of M1 station, the M-track sample 
selection starts from the T-track associated to a muon candidate 
from the $J/\psi$ and a cut on momentum of 12~GeV/c is applied.

\vspace*{5truemm}
\subsection{Background subtraction}
\label{sec:bg}

The presence of background clusters affects the search results. 
The necessary background subtraction requires different procedures for M2--M5 and for M1 station where the occupancy is higher. 
 
Assuming a Poissonian nature of the background, the average number of 
clusters due to the background and
the efficiency in a given search window can be extracted by a fit to
the  distribution of the number of clusters found.
If this method works, the background estimate in the neighbouring 
of the muon track takes 
automatically into account any possible correlation between  
muons and background, as in the case of delta rays or punch 
through in the calorimeter. 

For stations M2--M5, a fully satisfactory result is 
obtained assuming a background with two Poissonian components;
the probability of finding $n$ clusters in the search window 
being
\begin{equation}
  \label{eq:M2M5fit2}
P(n)= \epsilon \cdot [r \cdot \frac{ B_{1}^{n-1} \cdot e^{-B_{1}}}{(n-1)!} + (1-r)\cdot
\frac{B_{2}^{n-1} \cdot e^{-B_{2}}}{(n-1)!}]
+ (1-\epsilon) \cdot [r \cdot \frac{ B_{1}^{n} \cdot e^{-B_{1}}}{n!} + (1-r) \cdot 
\frac{B_{2}^{n} \cdot e^{-B_{2}}}{n!}]
\end{equation}
where  $\epsilon$ is the efficiency to be measured, B1 and B2 are the two Poissonian components of the background and $r$ their ratio.

\begin{figure}[h]
\begin{center}
\includegraphics[width=0.8\textwidth]{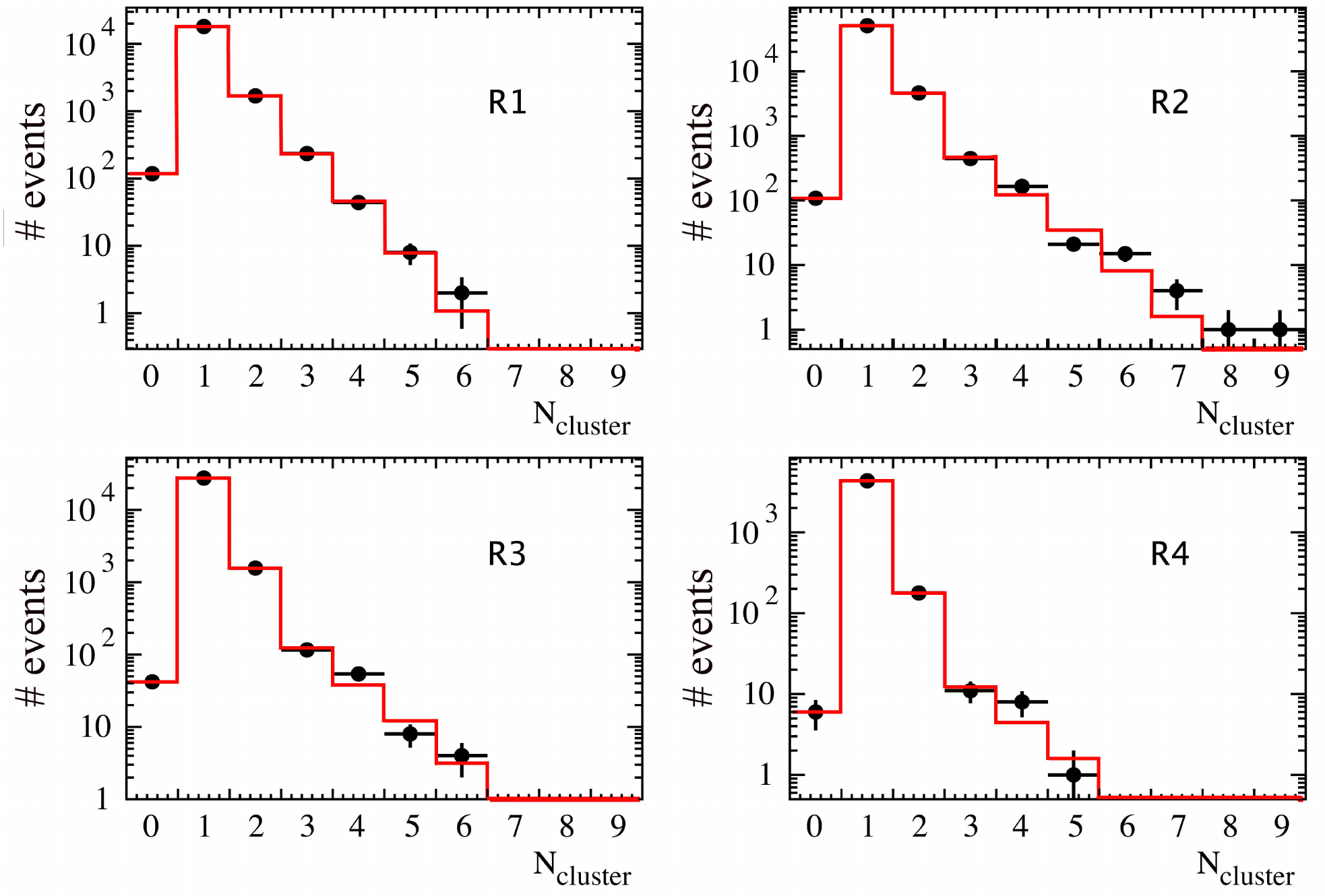}
\protect\caption{\small Multiplicity of clusters found in the
 8$\sigma$ search window for 
the four M3 regions. The thick line shows the results of the fit with the two components background described in equation \ref{eq:M2M5fit2}.}
\protect\label{fig:M3fit2}
\end{center}
\end{figure}

As an example, in Fig.~\ref{fig:M3fit2}
are reported the results of the fit in the case of an opening window of 
8$\sigma_x$ and 8$\sigma_y$ for the station M3.
It is worth to add that a fit with only one Poissonian component 
for the background does not give an equally good representation of the high multiplicity bins, but the fitted values of the efficiency do not show  significant differences.

\begin{figure}[h]
\begin{center}
\includegraphics[width=0.8\textwidth]{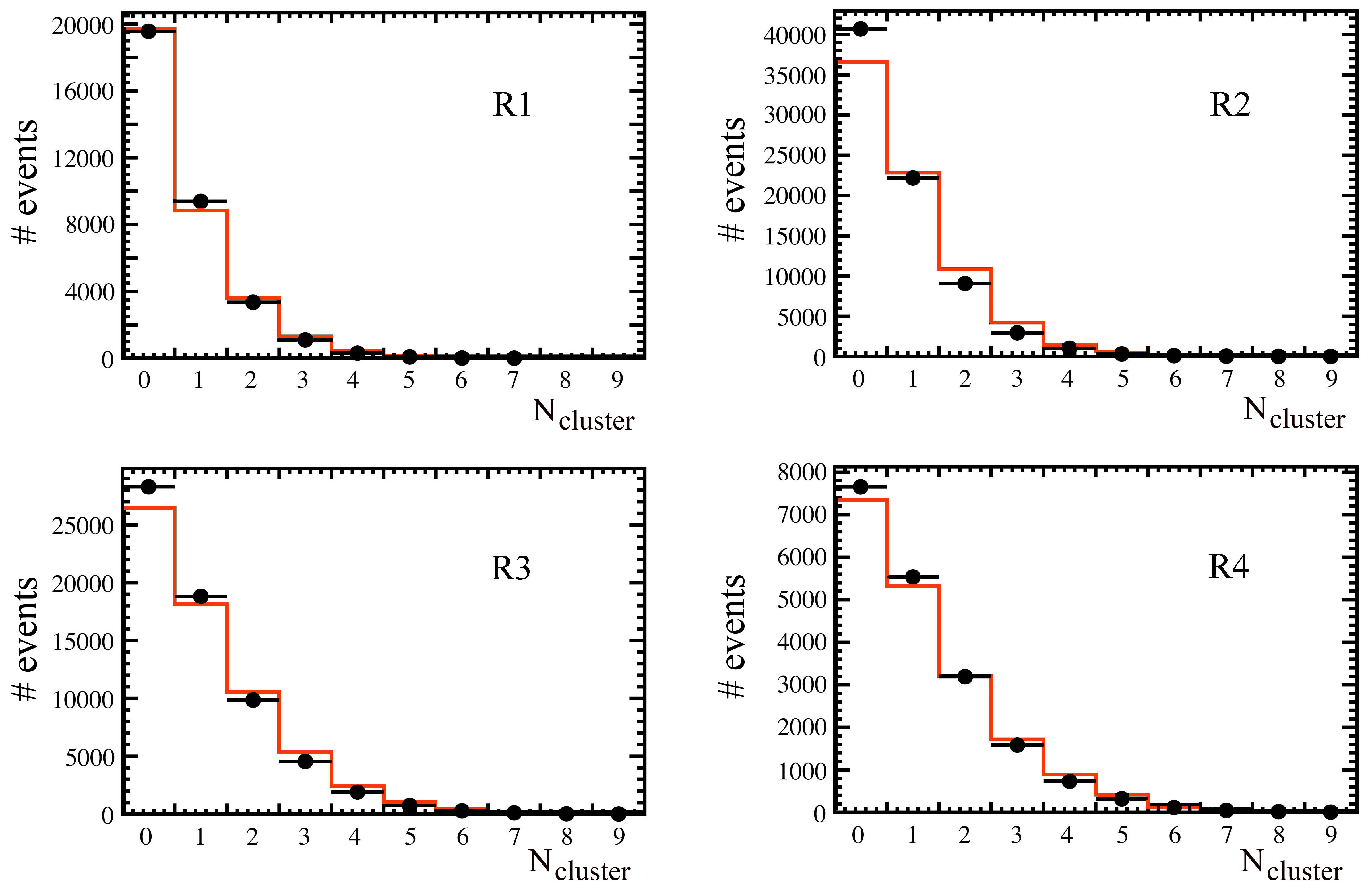}
\protect\caption{\small The cluster multiplicity in the 8$\sigma$ search window diminished by 1 
(histogram) and the cluster multiplicity in the corresponding  window in the opposite quadrant (dots), for $J/\psi$ muons in the four regions of M1 station.}
\protect\label{fig:M1sqvsoq}
\end{center}
\end{figure}

In the case of the more crowded station M1 where the $J/\psi$ 
muon sample is used,
the cluster multiplicities are not very well fitted assuming a background 
with two Poissonian components. 
Thus another fully independent method is also used to evaluate 
the background and the efficiency.
It exploits the  $\phi$ rotation invariance of the primary
interactions and assumes that the background correlated with the muon track is negligible.
The background is estimated by counting the average number of
clusters in the same search window but in the
opposite quadrant, with respect to the track prediction.
The soundness of the method is confirmed by comparing the number of
clusters found in the opposite quadrant with the number of clusters in the
track prediction quadrant having subtracted one cluster attributed to the muon track,
as shown in Fig.~\ref{fig:M1sqvsoq}.

The true efficiency $\epsilon_t$ is then estimated by the formula:
\begin{equation}
  \label{eq:M1effi}
  \epsilon_{a}=\epsilon_{t} +(1-\epsilon_{t})\cdot P_{bg}
\end{equation}
where $\epsilon_a$ is the apparent efficiency calculated as 
$N_{NCLUS>0} /N_{Preds}$, being  $N_{NCLUS>0}$ the number of 
tracks where at least one cluster has been found  
and $N_{Preds}$ the total number of tracks predicted to fall in 
the search window;  $P_{bg}=N_{NCLUS>0,OQ} /N_{Preds}$  is 
the probability to find at least one cluster in the search 
window in the opposite quadrant ($OQ$).

The efficiency values obtained with this
method do not show any significant difference with the values obtained with the previous one. 

\vspace*{5truemm}
\subsection{Check of the procedure with simulation}\label{sec:check}

To evaluate if the whole procedure of track selection, background 
subtraction and efficiency determination is robust and unbiased, 
a test has been made using events simulated with a Monte Carlo 
where all background components, like spillover hits and detector 
noise, are described and detector effects, such as cross-talk and 
chamber time jitter, are included.

The efficiency values computed on Monte Carlo events are well in agreement 
with the simulation parameters assumed and close to the results 
obtained with real data. However, to test the validity of the method, it 
is 
not so important to reproduce the absolute values found in data as to compare the efficiencies measured with Minimum Bias (for M2--M5) and $J/\psi$ (for M1) Monte Carlo events with the ones obtained applying the same procedure to another Monte Carlo sample of 
ideal muon tracks, the so-called 
{\em Particle Gun} (PG) muons. These muons are generated as starting from the interaction point 
with a predefined momentum and angular distribution.  
Since those events are practically background free and do not 
suffer from fakes in track reconstruction and selection, 
they allow to cleanly 
extract the detector efficiency with the present method. Then, independently of the absolute values found, a satisfactory agreement between the two samples would be an overall check of the correctness of the procedure.
The efficiency values were found to agree within the errors for the stations M1--M4, while 
for R3 and R4 regions of the M5 
station the 
efficiency extracted from the Minimum Bias is lower than the corresponding 
PG efficiency as shown in Tab.~\ref{tab:mbpg}.  
This difference is explained with a lack of purity of the Minimum Bias track sample: part of the selected tracks are muons 
produced in the showers inside the calorimeter that are aligned and matched with the hadron 
track initiating the shower. These muons have sometimes low momentum and are absorbed between M4 and M5. 
This effect is region dependent\footnote{Tracks crossing 
outer regions have, on average, lower 
momentum than those crossing inner regions.} and can cause an artificial inefficiency in M5.
Removing the muons with momentum below 6~GeV/c, the efficiencies reach  
values in agreement with the PG sample as can be seen in Tab.~\ref{tab:mbpg}.
To take into account  such effect, the real data results will be corrected for the ratio between the efficiencies 
with and without the 6~GeV/c cut, as extracted by the Minimum 
Bias Monte Carlo sample.

\begin{table}[h]
\begin{center}
\protect\caption{M5 Monte Carlo efficiency (\%) for Minimum Bias (L0MB) events with and without 6~GeV/c momentum cut compared  to {\em Particle Gun} (PG) muons. Statistical errors have been evaluated with an approximated binomial 68\% confidence interval.}
\vspace*{4truemm}
\protect\label{tab:mbpg}
\renewcommand{\arraystretch}{1.5}
\begin{tabular}{|c|c|c|c|c|} \hline
  & R1 & R2 & R3 & R4  \\ 
\hline 
PG&$99.50\;{^{+ \,0.14}_ {-\, 0.58}}$  &$99.73\;{^{+\, 0.04}_ {- \,0.16}}$  
&$99.69\;{^{+\, 0.03}_{-\, 0.09}}$  &$99.88\;{^{+ \,0.02}_{-\, 0.05}}$  \\
\hline
L0MB&$99.58\;{^{+ \,0.05}_{-\, 0.10}}$  &$99.46\;{^{+\, 0.04}_{-\, 0.05}}$  
&$99.14\;{^{+\, 0.06}_{-\, 0.08}}$  &$99.00\;{^{+\, 0.16}_{-\, 0.25}}$  \\
\hline
L0MB ($p>6$ GeV/c)&$99.77\;{^{+\, 0.03}_{-\, 0.12}}$  &$99.63\;{^{+\, 
0.03}_{- \,0.05}}$  
&$99.51\;{^{+\, 0.06}_{-\, 0.10}}$  &$99.80\;{^{+\, 0.02}_{-\, 0.22}}$  \\
\hline
\end{tabular}
\end{center}
\end{table}

\subsection{Measured efficiencies}\label{sec:results}

The behaviour of the efficiency as a function of the number
of $\sigma$'s of the opening window has been analysed for each
region of the muon system.
In all cases a correct saturation is observed at 3-4$\sigma$ 
demonstrating the reliability of the method. 
However the final value taken for the efficiency is the one at 8$\sigma$ to allow for the presence of non gaussian tails in the prediction point.

\begin{figure}[h]
\begin{center}
\includegraphics[width=0.75\textwidth]{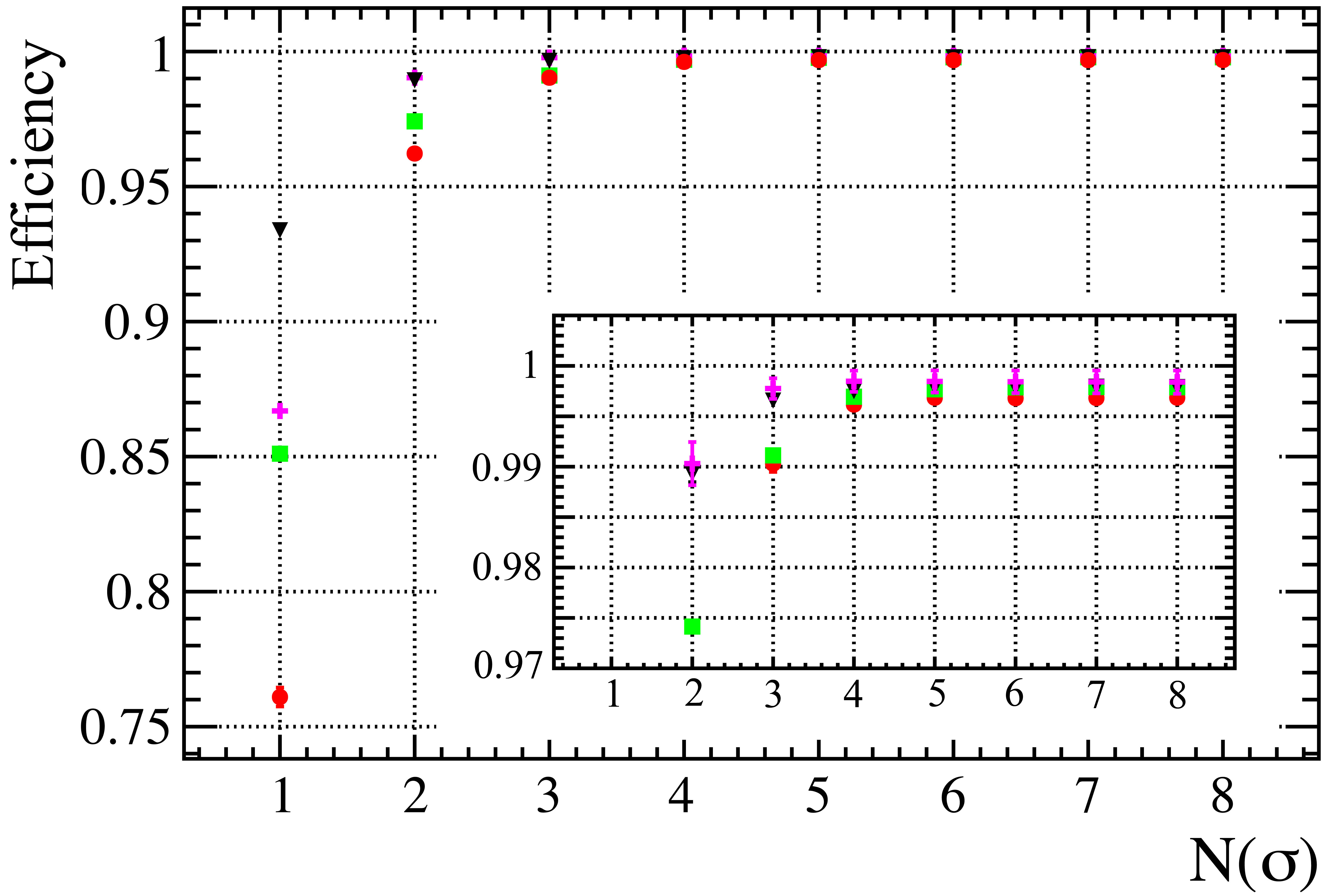}
\protect\caption{The measured efficiency of M2 station
as a function of the number of $\sigma$ of the search window.
The four regions are shown: R1 (red circles), R2 (green squares), R3 (black triangles), R4 (pink crosses). A zoom of the high efficiency region is shown in the insert.}
\protect\label{fig:saturam2}
\end{center}
\end{figure}

Fig.~\ref{fig:saturam2} shows
the behaviour of the efficiency as a function of the number of
$\sigma$ for the four regions of M2 station.

\begin{figure}[h]
\begin{center}
\includegraphics[width=0.75\textwidth]{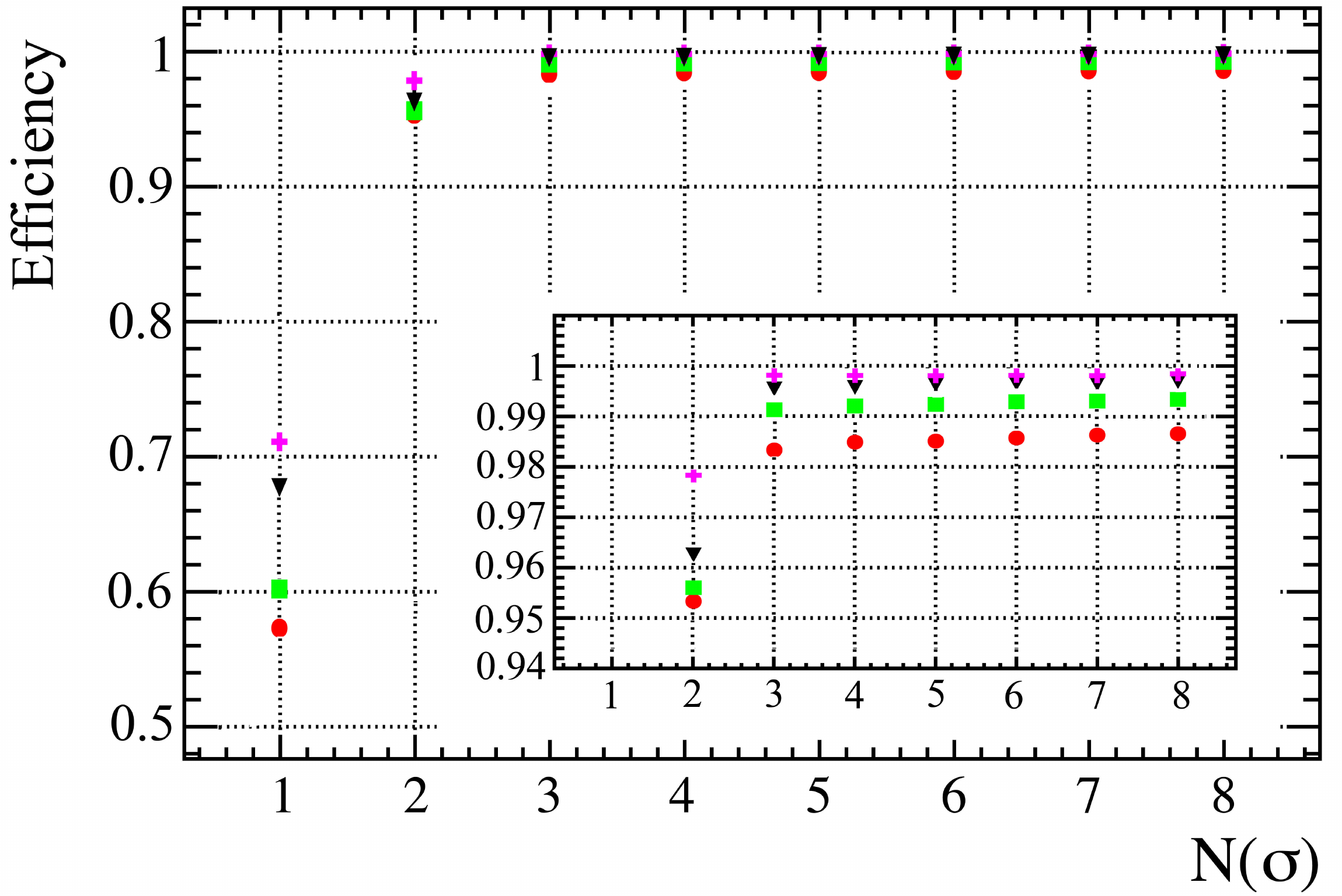}
\protect\caption{The measured efficiency of M1 station
as a function of the number of $\sigma$ of the search window.
The four regions are shown:
 R1 (red circles), R2 (green squares), R3 (black triangles), R4 (pink crosses). A zoom of the high efficiency region is shown in the insert.}
\protect\label{fig:saturam1}
\end{center}
\end{figure}

In Fig.~\ref{fig:saturam1} the same quantity is shown
for the station M1, with the efficiencies obtained with 
equation \ref{eq:M1effi}.

The efficiency for M1 was at the end evaluated as the average of the efficiencies obtained with the two methods corresponding to equations
\ref{eq:M2M5fit2} and \ref{eq:M1effi}, which turn out to be in excellent agreement.

\par In Tab.~\ref{tab:Summary}
the measured efficiencies with the statistical and
systematic errors are reported.
For M2--M5 the systematic errors due to background modeling
have been estimated by changing the fit function from two poissonians
to a single one. The efficiency varies less than 0.01\% in all regions of M2--M5.
For M1 where two independent methods of estimating the background have been used,
a systematic uncertainty of half the
difference between the two results has been assumed.
Since the choice of evaluating the final efficiency at 8$\sigma$
has a certain degree  of arbitrariness, a systematic error
of half the difference between the efficiency value
calculated at 4$\sigma$ and 8$\sigma$ has been assumed.
The uncertainty, due to MC limited statistics, of the correction applied 
on the M5 efficiency to take  into account the absorption
of muons between M4 and M5, is included in the systematics.
The different sources give comparable systematic uncertainties that have been added in quadrature.

\begin{table}[h]
\begin{center}
\protect\caption{ Efficiency values (\%) of all the twenty regions with their statistical and systematic errors. Statistical errors have been evaluated with an approximated binomial 68\% confidence interval. Systematic errors are calculated as the sum in
quadrature of various  contributions - see text for details.
}
\vspace*{5truemm}
\protect\label{tab:Summary}
\renewcommand{\arraystretch}{1.5}
\begin{tabular}{|c|c|c|c|c|} \hline
 &  R1 & R2 & R3 & R4  \\
\hline
M1& $ 98.66^{+0.07}_{-0.07}\pm 0.08 $ & $ 99.37^{+0.04}_{-0.04}\pm 0.08 $ & $ 99.70^{+0.02}_{-0.04} \pm  0.06$  & $99.85^{+0.01}_{-0.08}\pm 0.02$\\  
\hline
M2& $ 99.68^{+0.03}_{-0.07}\pm 0.04 $ & $99.78^{+0.02}_{-0.04} \pm 0.05$ &$ 99.79^{+0.02}_{-0.06}\pm 0.02 $ & $99.80^{+0.02}_{-0.22}\pm 0.01$\\  
\hline
M3&$99.35^{+0.05}_{-0.04} \pm 0.05 $ &$ 99.79^{+0.02}_{-0.02} \pm 0.04 $ & $99.85^{+0.01}_{-0.03}\pm 0.01 $ & $ 99.86^{+0.01}_{-0.14} \pm 0.01$\\  
\hline
M4&$ 99.62^{+0.03}_{-0.09} \pm 0.07  $& $ 99.89^{+0.01}_{-0.01} \pm 0.03 $ & $ 99.65^{+0.03}_{-0.05} \pm 0.03 $ & $ 99.72^{+0.03}_{-0.17} \pm 0.01$\\  
\hline
M5&$ 99.64^{+0.05}_{-0.07} \pm 0.10 $ & $ 99.82^{+0.02}_{-0.04} \pm 0.04  $& $ 99.90^{+0.02}_{-0.07} \pm 0.12  $ & $ 100.0^{+0.00}_{-0.46} \pm 0.1$\\  
\hline
\end{tabular}
\vspace*{6truemm}
\end{center}
\end{table}

\begin{figure}[htb]
\begin{center} 
\mbox{\includegraphics[width=.75\textwidth]{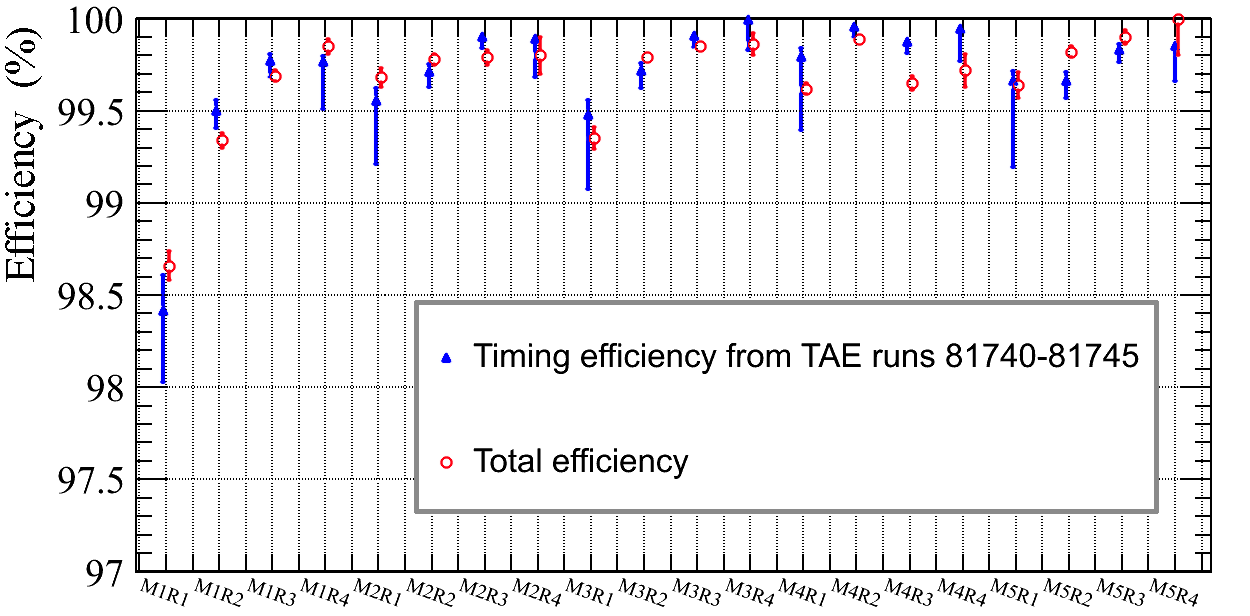}}  
\protect\caption{The results for total efficiency are compared with
  the estimates of timing efficiency from section
  \ref{sec:timingresults}. Errors are statistical only. For the
  regions more affected by combinatorial background,
  a systematic underestimation of the timing efficiency by a few per
  mill is present.} 
\protect\label{fig:cmpEffs}
\end{center}
\end{figure}

The overall detector efficiency is
compared in Fig.~\ref{fig:cmpEffs} with the results of section~\ref{sec:timingresults} where the
contributions of timing tails to the inefficiency are evaluated. As already
mentioned, the limited statistics of TAE data prevented to apply the
same track selection for the two studies. The lower
purity of the muon track sample for the timing analysis leads to a slight
underestimate of the timing efficiency. 
It can be added that the TAE sample was also acquired at a different time, at the end of the runs, with a lower high voltage in the region M1R1 (see section~\ref{sec:functioning}) and could show a
poorer performance. 
Nevertheless, the region dependence of the two
measurements is very similar in the errors, and for regions less affected by combinatorial background the timing efficiency is found to be
compatible or only slightly better than the overall efficiency. This indicates
that the bulk of the detector inefficiency comes from signals 
falling outside the 25~ns LHC gate and the intrinsic chamber inefficiencies, or the small geometrical losses give minor contributions.

